\documentclass[iop,twocolumn]{emulateapj}
\usepackage{amsmath,amssymb,amsfonts,bm}
\usepackage{graphicx}
\usepackage{graphics}
\usepackage{keyval}
\usepackage{trig}
\usepackage{float}
\usepackage[dvipsnames]{color}
\usepackage{calc}
\usepackage{hyperref}

\def\beq{\begin{equation}}
\def\eeq{\end{equation}}
\def\bey{\begin{eqnarray}}
\def\eey{\end{eqnarray}}

\def\pc{\, {\rm pc} }
\def\kpc{\, {\rm kpc} }
\def\mpc{\, {\rm Mpc} }

\def\sun{\odot}
\def\Msun{M_\odot}

\def\kms{\, {\rm km \, s}^{-1} }

\def\a0{$a_0$}

\def\ixxt{{\tilde I_{\rm xx}}(r)}

\def\ixyt{{\tilde I_{\rm xy}}(r)}

\def\ixxtnew{{\tilde I_{\rm x'x'}}(r)}
\def\iyytnew{{\tilde I_{\rm y'y'}}(r)}
\def\izztnew{{\tilde I_{\rm z'z'}}(r)}
\def\at{{\tilde a}}
\def\bt{{\tilde b}}
\def\ct{{\tilde c}}
\def\pr{{\tilde p}_{\rm 90\%}}
\def\qr{{\tilde q}_{\rm 90\%}}
\def\atr{{\tilde a}_{\rm 90\%}}
\def\btr{{\tilde b}_{\rm 90\%}}
\def\ctr{{\tilde c}_{\rm 90\%}}
\def\tdyn{t_{\rm dyn}}

\begin{document}
\title{Lopsidedness of self-consistent galaxies by the external field effect of clusters}
\author{Xufen Wu}\email{xufenwu@ustc.edu.cn}
\affiliation{CAS Key Laboratory for Research in Galaxies and Cosmology, Department of Astronomy,\\
  University of Science and Technology of China, Hefei, 230026, P.R. China\\
  School of Astronomy and Space Science, University of Science and Technology of China, Hefei 230026, P.R. China\\
School of Physics and Astronomy, University of St Andrews, North Haugh, Fife, KY16 9SS, UK}
\author{Yougang Wang}
\affiliation{Key Laboratory of Computational Astrophysics, National Astronomical Observatories,\\
Chinese Academy of Sciences, Beijing, 100012, P.R. China}
\author{Martin Feix}
\affiliation{CNRS, UMR 7095 \& UPMC, Institut d'Astrophysique de Paris, 98 bis Boulevard Arago, 75014 Paris, France\\
Department of Physics, Israel Institute of Technology - Technion, Haifa 32000, Israel
}
\author{HongSheng Zhao}
\affiliation{School of Physics and Astronomy, University of St Andrews, North Haugh, Fife, KY16 9SS, UK\\
Key Laboratory of Computational Astrophysics, National Astronomical Observatories,\\
Chinese Academy of Sciences, Beijing, 100012, P.R. China}

\begin{abstract}
Adopting Schwarzschild's orbit-superposition technique, we construct a series of self-consistent galaxy models, embedded in the external field of galaxy clusters in the framework of Milgrom's MOdified Newtonian
Dynamics. These models represent relatively massive ellipticals with a Hernquist radial profile at various distances from the cluster centre.
Using $N$-body simulations, we perform a first analysis of these models and their evolution. We find that self-gravitating axisymmetric density models, even under a weak external
field, lose their symmetry by instability and generally evolve to triaxial configurations. A kinematic analysis suggests that the instability originates from both box and
non-classified orbits with low angular momentum. We also consider a self-consistent isolated system which is then placed in a strong external field and allowed to evolve freely.
This model, just as the corresponding equilibrium model in the same external field, eventually settles to a triaxial equilibrium as well, but has a higher velocity radial anisotropy
and is rounder. The presence of an external field in MOND universe generically predicts some lopsidedness of galaxy shapes.
\end{abstract}

\maketitle

\section {Introduction}
Elliptical galaxies widely exist in the Universe, either isolated or embedded in clusters of galaxies, and they have compact centres. Studies of their dynamical behaviour and evolution
typically require the building of equilibrium models, which is quite challenging for elongated or triaxial systems. Schwarzschild's orbit-superposition method \citep{Schwarzschild1979,
Schwarzschild1982} is a powerful technique to find self-consistent solutions for spherical, axisymmetric and triaxial systems. Adopting Schwarzschild's method, the aim of this work is
to construct such equilibrium models for elliptical galaxies in external fields, i.e. galaxies that are gravitationally bound within clusters, using the framework of Milgrom's MOdified
Newtonian Dynamics \citep[MOND;][]{Milgrom1983c, BM1984}. In addition, the kinematic properties and stability of these systems are explored with the help of $N$-body simulations.

The approach of Schwarzschild can be divided into three basic steps:
\begin{enumerate}
\item An analytic density distribution is chosen and the corresponding gravitational potential is calculated.
The whole system is then segmented into many equal mass cells.
\item A full library of orbits within the previously calculated potential is computed, and the time spent in each cell is recorded.
\item The non-negative linear superposition of orbits which reproduces the original density profile is determined.
\end{enumerate}

The method of Schwarzschild has been applied to test various density models for self-consistency, including pattern-rotating barred galaxies \citep{Zhao1996_bar,Wang_etal2012,Wang_etal2013}. Many early applications of Schwarzschild's approach assumed constant density cores. E.g., \cite{Statler1987} found self-consistency of the perfect ellipsoid models of \cite{deZeeuw_Lynden-Bell1985}. However, observations showed that almost all elliptical galaxies have central densities that follow a power law $\rho\sim r^{-\gamma}$ \citep{Moller_etal1995, Crane_etal1995, Ferrarese_etal1994, Lauer_etal1995}. Low-luminosity ellipticals
have steeper centres, $\gamma\approx 2$, while the most luminous ellipticals have shallower ones, $0\leq\gamma\leq 1$. Also, \citet{Tremblay_Merritt1996} found that the intrinsic shapes
of ellipticals depend on their luminosity: the short-long axis ratio of the most luminous ellipticals has a peak at 0.75 whereas that of low-luminosity ellipticals peaks at 0.65.
\cite{dehnen1993} proposed a family of models whose density distributions follow $\rho\sim r^{-\gamma}$ in the central region and $\rho \sim r^{-4}$ at distant radii, where $\gamma$
is a free parameter. Various studies \citep{Merritt_Fridman1996, Rix_etal1997, Poon_Merritt2004, Capuzzo-Dolcetta_etal2007} of Schwarzschild's technique applied to triaxial Dehnen
profiles have been conducted, and it has been shown that self-consistent solutions for these models can be constructed in the case of Newtonian gravity.

As an alternative to cold dark matter on galactic scales, the MOND paradigm is built on the tight relation between the distribution of baryons and the gravitational field in spiral
galaxies \citep{McGaugh_etal2007, FGBZ2007}. In fact, its simple formulation leads to excellent predictions of rotation curves for galaxies ranging over five decades in mass (see,
e.g., \citealt{spiral1, mondref1}), including our own Milky Way \citep{FB2005, FBZ2007}. The successes and problems of MOND are extensively discussed in \cite{Famaey_McGaugh2012}.
Moreover, MOND has recently been used to explain the rotational speed in polar rings \citep{Lughausen+2013}, the formation of shell structure in the elliptical galaxy NGC 3923
\citep{Bilek_etal2013,Bilek_etal2014}, the velocity dispersion of Andromeda dwarf galaxies \citep{McGaugh_Milgrom2013}, and the mass discrepancy-acceleration correlation of disc
galaxies \citep{Milgrom1983a, Sanders1990, McGaugh2004,Wu_Kroupa2015} and of pressure-supported systems \citep{Scarpa2006}. It also provides constraints on the mass-to-light ratio
derived from the vertical stellar velocity dispersion \citep{Angus_etal2016}. 

In contrast to the Newtonian case, the internal dynamics of a gravitating system in MOND is affected by any external fields, i.e. even a freely falling system in MOND will exhibit
a dynamical evolution different from that of an isolated one. This attribute implies a violation of the strong equivalence principle and is usually referred to as the external field
effect \citep{Milgrom1983a}. The impact of external fields has been studied for a variety of different situations, including the motion of probes in the inner solar system
\citep{Milgrom2009}, the Roche lobe of binary stars \citep{ZT2006}, the kinetics of stars in globular clusters \citep{Milgrom1983a, HVSs}, the escape speeds and truncations of
galactic rotation curves \citep{Wu_etal2007}, satellites surrounding a host galaxy \citep{BM2000, Tiret_etal2007, Angus2008}, and the phase transition of distant star clusters moving
towards the galactic centre \citep{Wu_Kroupa2013}.

Schwarzschild's orbit-superposition method has already been applied within Milgromian dynamics: for example, models of ellipsoidal field galaxies were found both self-consistent
and stable \citep{triaxial, stability}. Further, the morphology of elliptical cluster galaxies was discussed by \cite{Wu_etal2010} and exhibits lopsided shapes along the external
field direction. In what follows, we will mainly focus on the kinematic aspects of these models and the numerical details of our approach, extending the analysis to triaxial systems
in a gravitating environment. Although it yields a less realistic scenario, we only consider external fields which are uniform and constant. Independent of the framework, external
tidal fields will influence a system's internal dynamics and may obscure fundamental differences between gravity theories. To maximally distinguish between Newtonian dynamics and
MOND, we thus ignore tidal effects in our analysis.\footnote{Generally, it is desirable to have a full treatment of the problem including tidal effects. This would allow to explore
other, more complex scenarios such as evolution in fast-varying backgrounds, e.g. a galaxy crossing the centre of cluster, and will be subject to future work.} Such an idealised
case corresponds to systems either entirely restricted within the tidal radius or moving in a smooth and slowly varying background field, for example, a galaxy circularly orbiting
the cluster centre. In these situations, an external field is mainly dominated by its uniform part, and tidal effects play only a subordinate role.

Posing already a challenge in Newtonian gravity, the construction of equilibrium models for elongated or triaxial systems in MOND is additionally hampered by the nonlinear modification
of Poisson's equation. This is especially true for systems with external fields because their phase-space distribution is determined by both internal density and external field. In their
galaxy merger simulations, \citet{NLC2007b} obtained the distribution function of a Hernquist sphere by Eddington inversion in a MONDian potential. For more complex triaxial models,
however, it is generally not possible to obtain analytic solutions to Eddington's equation and this procedure becomes very difficult. \cite{TC2007,TC2008} employed Newtonian equilibrium
models for spiral galaxies embedded into a Plummer-type dark halo, and replaced gravity by MONDian dynamics afterwards. Similarly, the simulations by \cite{Haghi_etal2009} made use of
Newtonian models initially, but then particle velocities were increased to avoid gravitational collapse of globular cluster models in the Milky Way. If set up in this way, such initial
conditions will immediately cause the system in question to relax until it reaches a new state of equilibrium. This will remain true when using Schwarzschild's approach which was designed
for genuine equilibrium systems. For example, initially axisymmetric models will start developing asymmetric shapes \citep{Wu_etal2010}. Below we shall investigate this relaxation process
in more detail, including both axisymmetric and triaxial configurations.

The stability of disk galaxies hosted by dark matter halos in Newtonian gravity has been studied by \citet{Sellwood_Evans2001}. It has been shown that the lopsided instability ($m=1$
mode) can be avoided when their massive outer disks are tapered, and the galaxies are stabilised by dense centers. Further, \citet{DeRijcke_Debattista2004} found off-centered nuclei
in flattened non-rotating systems, and a promising mechanism is the destruction of box orbits. The growth of two different $m=1$ modes, associated with a Jeans-type instability for
counter-rotating disk models and swing amplification for fully rotating disk models, was studied by \cite{Dury_etal2008}. As most observed galaxies appear lopsided \citep[$m=1$
perturbations;][]{Rix_Zaritsky1995,Haynes_etal1998,deZeeuw_etal2002}, this motivates investigating such mechanisms also in the context of MOND. From an analytic point of view, there
is still little known about the stability of galaxies in MOND \citep{nbody,stability,Nipoti_etal2011}, and, in particular, there exist no stability studies in external or tidal fields.
A numerical study thus provides a first step into this direction. 

The paper is organised as follows: In \S\ref{model}, we introduce our basic setup and discuss the effect of external fields on static galaxy potentials. In \S\ref{schwarzschild}, we
use Schwarzschild's technique to construct quasi-equilibrium models of galaxies, and perform a kinematic analysis of these systems in \S\ref {nbody}. Finally, we compare our results
to the evolution of isolated galaxies, and conclude in \S\ref{concl}.

\section{Mass models and static potentials}
\label{model}
\subsection{Density profiles and external fields}
The MONDian Poisson's equation including the presence of an external gravitational acceleration $\vec{g}_{\rm ext}$ reads \citep{Wu_etal2007}
\beq
\vec\nabla\cdot\left\lbrack\mu\left (\frac{|\vec{g}|}{a_{0}}\right)\vec{g}\right\rbrack = 4\pi G\rho_{b},\quad
\vec{g} = \vec\nabla\Phi_{\rm int} + \vec{g}_{\rm ext}.
\eeq
Here $\Phi_{\rm int}$ is the internal potential generated by the baryon density $\rho_{b}$ and $a_{0}\approx 3700~\rm {km}^2 \rm s^{-2}\kpc^{-1}$ is Milgrom's
constant. To produce both a Newtonian and MONDian limit, the interpolating function $\mu(x)$ has to be of the following form $(x=\vec{g}/a_{0})$:
\beq
\begin{split}
\mu(x) &\sim x \qquad x \ll 1,\\
\mu(x) &\sim 1 \qquad x \gg 1.
\end{split}
\eeq
In what follows, we will use the simple form of the $\mu$-function adopted by \citep{FB2005}, which is
\beq\label{simplemu}
\mu(x)={x\over 1+x}.
\eeq
Although observationally excluded in the Solar System, this $\mu$-function is still a viable choice on galactic scales and in good agreement with
the terminal velocities of the Milky Way and NGC3198. Compared to the ``standard'' form introduced in \cite{Milgrom1983c}, the transition between Newtonian gravity
and MOND happens more gradually.
For the baryonic density, we adopt a Hernquist profile \citep{hernquist},
\beq\label{den}
\rho (r) = {M \over 2\pi abc}{1\over r(1+r)^{3}},
\eeq
where
\beq
r = \sqrt{\left(x\over a\right)^2+\left(y\over b\right)^2+\left(z\over c\right)^2},
\eeq
and the constants $a,b,c$ are the typical length scales of the galaxy's major, intermediate and minor axes, respectively. Here we consider five different galaxy models:
four axisymmetric elliptical galaxies without and with external fields, respectively, and a triaxial galaxy model embedded in an external field. The parameters of the
galaxy models are listed in Table~\ref{mass}. These models represent medium-sized elliptical galaxies with masses on the order of $10^{10}$--$10^{11}\Msun$, bright enough
to observe the outer parts. In this case, the internal accelerations are comparable to several $a_0$. Because elliptical galaxies should lie on the fundamental plane
\citep{Djorgovski_Davis1987}, there is a strong correlation between stellar masses $M$ and effective radii $R_e$ \citep[Figure 13 of][]{Gadotti2009}. The effective radii
of galaxies with a total mass of $10^{10}$--$10^{11}\Msun$ range from $0.5$--$3\kpc$. For all five models, we set $a=1\kpc$ which lies within the $R_e$ dispersion range
of observational data points \citep{Gadotti2009}. 

The strengths of the external fields are chosen as $g_{\rm ext}=0.01a_0, ~0.1a_0$ and $1.0a_0$ for the axisymmetric models and $1.0a_0$ for the triaxial model, corresponding
to weak, intermediate and strong external fields, respectively. Hence for the strong external field cases, the internal and external accelerations at several $R_e$ (about
$10\kpc$ to the galactic centre) are comparable to each other. For the intermediate case (model $3$), the two accelerations become comparable at approximately $70 \kpc$, whereas
for the weak external field case (model $2$), this requires moving to very large radii, beyond hundreds of kpc.

\subsection{Distortion of static potentials in external fields}
To compute the static potential, we make use of the MONDian $N$-body solver NMODY \citep{CLN2006,NLC2007b}. NMODY is a particle-mesh code that assigns particles
by cloud-in-cell, and solves the modified Poisson's equation on a spherical grid, using a second-order leap-frog scheme for time integration. 
We adopt a resolution of
$n_r\times n_{\theta} \times n_{\phi} =512\times 128\times 256$ on a spherical grid, where the grid segments are defined as
\beq
\begin{split}
r_{i} &= 2\tan\left\lbrack(i+0.5)0.5\pi /(n_r+1)\right\rbrack\kpc,\\
\theta_{j} &= \pi \times (j+0.5)/ n_{\theta},\\
\phi_{k} &= 2\pi \times k / n_{\phi},
\end{split}
\eeq
where $i=0$ .. $n_r$, $j=0$ .. $n_{\theta}-1$, and $k=0$ .. $n_{\phi}-1$. 

\begin{figure}{}
\begin{center}
\resizebox{4.2cm}{!}{\includegraphics{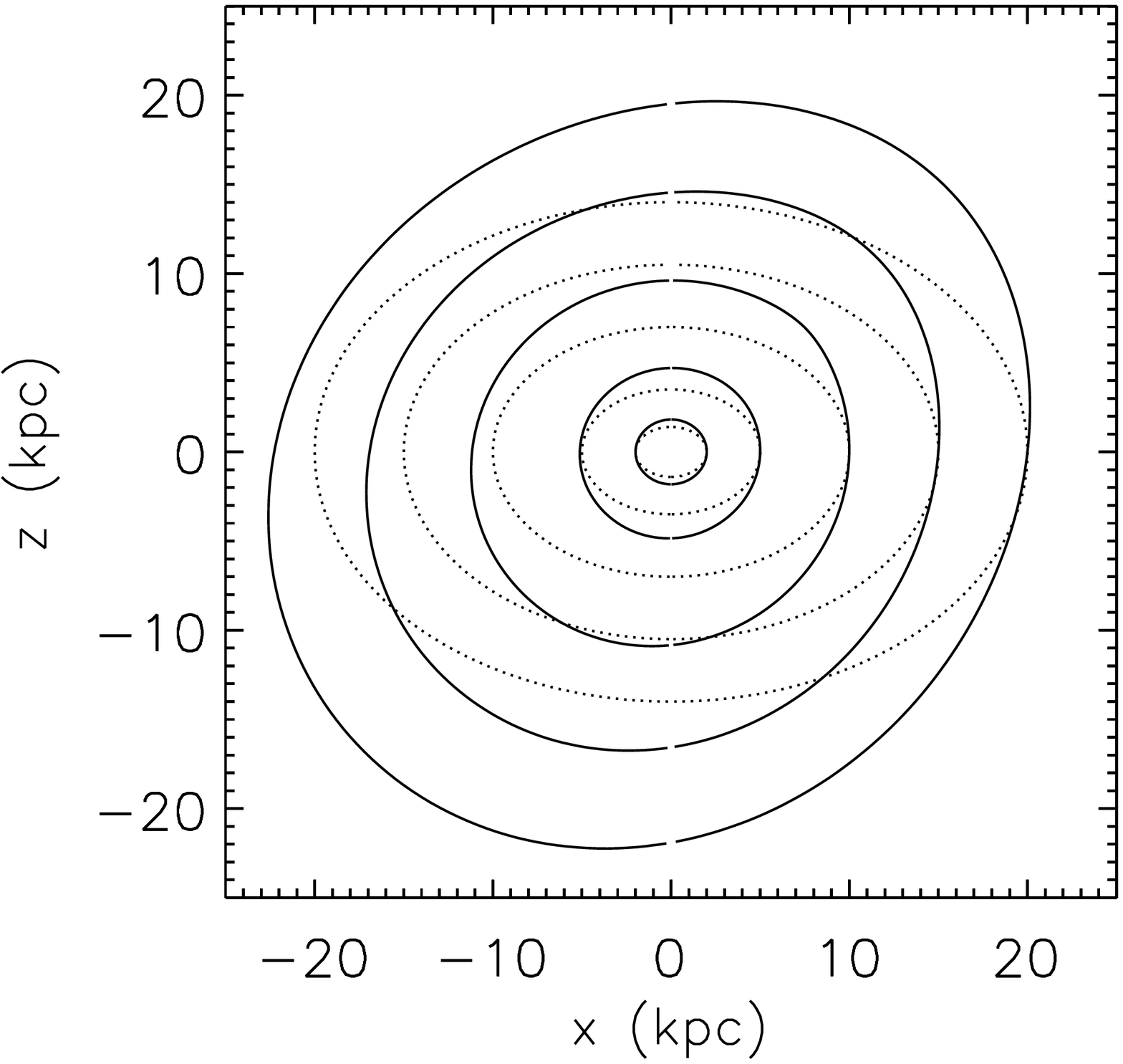}}\resizebox{4.2cm}{!}{\includegraphics{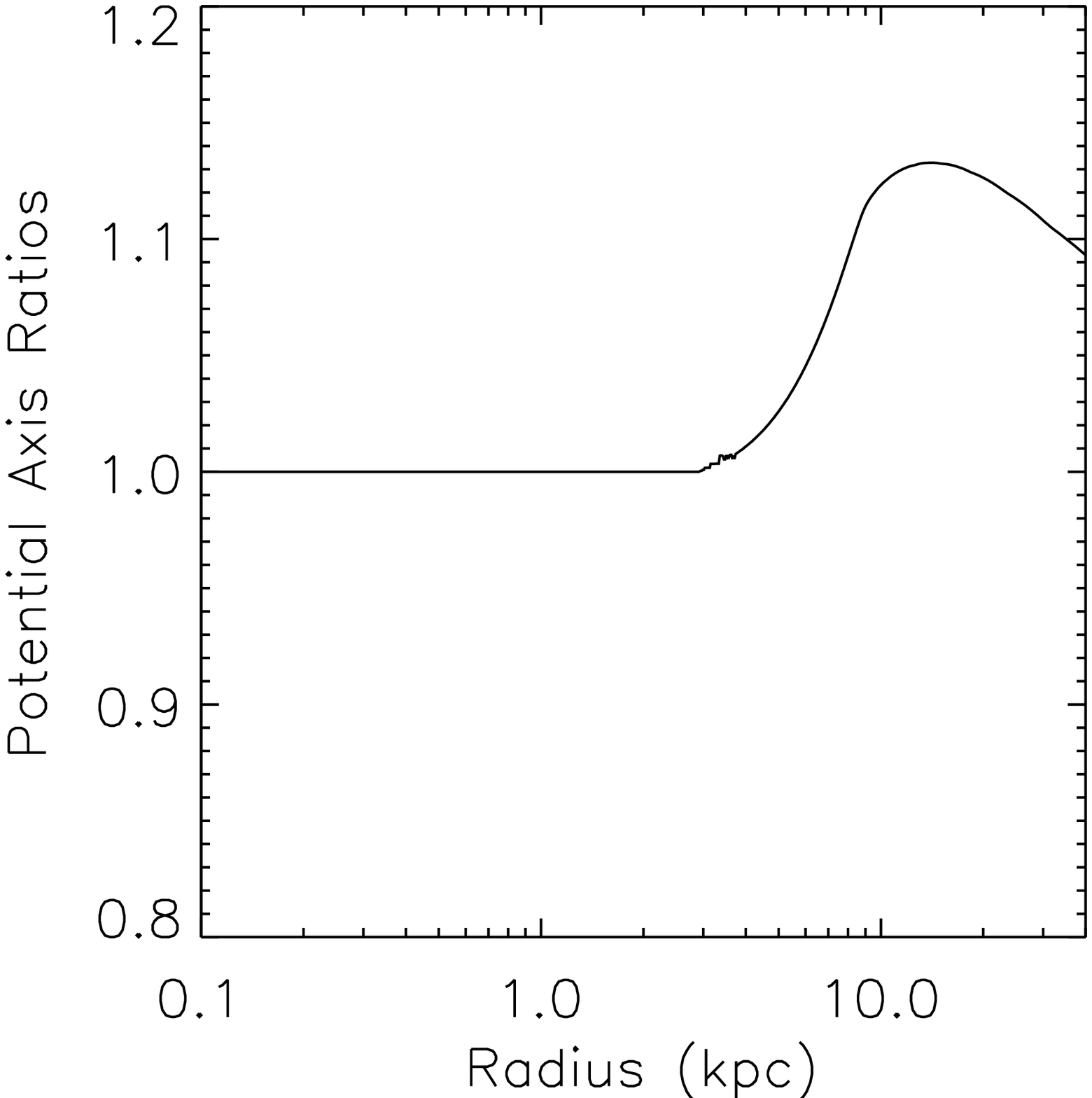}}\vskip 0.5cm
\makeatletter\def\@captype{figure}\makeatother
\caption{\textbf{Left panel}: Isodensity (dotted) and isopotential (solid) contours of model $4$. The isodensity contours correspond to ellipsoidal radii of
$2, 5, 10, 15$, and $20\kpc$. The isopotential contours are located at the same radii along major axis. The direction of the external field
is along the negative $x$-$z$ diagonal. \textbf{Right panel}: The ratio of isopotentials, $(r_{-} : r_{+})_{\Phi({r_+})}$, evaluated at the same radius in
anti-parallel and parallel directions of the external field for model $4$.}
\label{isopot}
\end{center}
\end{figure}

\cite{LMC} and \cite{Wu_etal2010} showed that the internal potentials are prolate with respect to the direction of the external field. This effect becomes
important when the internal and external fields are roughly of the same order; an even more significant effect occurs at weaker accelerations, i.e. in external
field dominated regions where $|\vec{g}_{\rm int}| << |\vec{g}_{\rm ext}| << a_0$.
In this case, one finds that the potentials are not only prolate, but also appear distorted. Figure~\ref{isopot} shows isodensity and isopotential contours
for model $4$ as listed in Table~\ref{mass}. The contours correspond to radii of $2, 5, 10, 15$, and $20\kpc$ along the major axis. We can easily identify a
distortion of the potential along the direction of the external field at large radii, $r \sim 10\kpc$, where internal and external field are comparable. The
density contours, however, are still axisymmetric. Consequently, models constructed with external fields may not be necessarily self-consistent. Considering
that the radius where the strongest external field (model $4$) starts dominating the internal dynamics is at about $10~\kpc$ (enlcosing $\approx 82\%$ of the
total mass), the models should not be significantly affected for the most part and reside in a quasi-equilibrium. Thus we expect Schwarzschild's method to be
applicable in these cases.

For a simplified view, let us consider a spherically symmetric system embedded into an external field. Integrating the MOND Poisson's equation, we arrive at
the following expression \citep{BM1984}:
\beq
\mu\left ({|\vec{g}|\over a_0}\right ) \vec{g} = \vec{g}_N + \vec\nabla\times\vec{h},
\label{integmond}
\eeq
where $g_N$ is the Newtonian gravitational field, and $\vec\nabla\times\vec{h}$ is a solenoidal vector field determined by the condition that $\vec{g}$ can be
expressed as the gradient of a scalar potential. Restricting ourselves to the axis parallel to the external field's direction, $\vec{g}$ and $\vec{g}_{N}$ must
be either parallel or anti-parallel, assuming that the symmetry centre coincides with the coordinate origin. Hence the curl term $\vec\nabla\times\vec{h}$ vanishes.
The strength of the total gravitational acceleration $g=|\vec{g}|$ along the negative semi-axis in the external field's direction, i.e. where the external field
cancels part of the internal one, is $g^-=|g_{\rm ext}-g_{\rm int}|$ while on the positive semi-axis it is $g^+=g_{\rm int}+g_{\rm ext}$. The two different sides
have different values of the $\mu$-function at the same radii, leading to a larger MONDian enhancement of gravitation along the negative semi-axis. Clearly, for
the underlying spherically symmetric density distribution, the potential and its derivatives are axisymmetric. Applied to a typical triaxial system, however, the
result is approximately the same. For an external field pointing into an arbitrary direction, such a system has no symmetries anymore, but the curl term in Eq.
\ref{integmond} only accounts for corrections on the level of $10\%$ \citep{Brada_Milgrom1995}.

Returning to the axisymmetric model $4$, Fig.~\ref{isopot} confirms the above considerations. On the \textbf{left panel}, the isopotential contours are denser in
the first octant than in the third one. The internal potential is shallower in the first octant where $g^+=|g_{\rm ext}+g_{\rm int}|$, and steeper in the third
octant where $g^-=g_{\rm int}-g_{\rm ext}$. As can be seen from the \textbf{right panel} of Fig.~\ref{isopot}, the lopsidedness of the potential reaches its maximum
at roughly $10\kpc$ where the external and internal fields are comparable to each other. The semi $x$-axis ratio $r^-:r^+$ of the isopotential contours reaches about
$1.14$ between $10$ and $11\kpc$. At small radii ($r<3\kpc$), the internal gravitational field dominates, and thus $r^-:r^+$ is basically $1$. At much larger radii
($r>>10\kpc$), the relative contribution of the external field increases, and $r^-:r^+$ falls down to $1$ again because the $\mu$-function approaches a constant,
$\mu = \mu(|\vec{g}_{\rm ext}|/a_0)$.

\section{Schwarzschild Technique and Model Self-consistency}\label{schwarzschild}

\begin{table*}
\begin{center}\vskip 0.00cm
\caption{Galaxy models and Schwarzschild parameters: The total mass $M$ of each model is expressed in units of $10^{10}\Msun$, $a,~b$ and $c$
are scale lengths in units of $\kpc$, and $g_{\rm ext}$ denotes the strength of the external field in units of $a_0\approx 3700 (\kms)^2\kpc$.
As the external field breaks the symmetry of the potential, the symmetry axes of potential and density differ from each other. The starting
octants refer to the symmetry plane of the potential. Here $N_{\rm cell}$ is the number of cells to impose self-consistency, $N_{\rm stationary}$
is the number of stationary starting orbits built in the orbital library, and $N_{\rm ejecting}$ is the number of ejecting orbits starting from
the $x$-$z$ plane. The total number of orbits is given by $N_{\rm orbit}$.}
\begin{tabular}{lllllc}
\\
\hline
Model &1 & 2 & 3 & 4 & 5 \\
\hline
M [$10^{10}\Msun$] & 5 & 5 & 5 & 5 & 10 \\
$a:b:c$ & 1: 1: 0.7 & 1: 1: 0.7 & 1: 1: 0.7 & 1: 1: 0.7 & 1: 0.86: 0.7\\
$g_{\rm ext}$ [$a_0$] & 0 & 0.01 & 0.1 & 1.0 & 1.0 \\
Direction of $\vec{g}_{\rm ext}$ & - & negative $z$-$x$ & negative $z$-$x$ & negative $x$-$z$  &  negative\\
  &  & diagonal & diagonal & diagonal & $x$-axis \\
Density symmetry & axisym & axisym & axisym & axisym & triaxial\\
Potential sym. Axes & $x$,$y$,$z$ & $y$ & $y$ & $y$ & $y$,$z$ \\
Starting octants & $y$ & I,II,V,VI & I,II,V,VI & I,II,V,VI & I, II \\
Reflecting planes & $x$-$y$, $x$-$z$, $y$-$z$ & $x$-$z$ & $x$-$z$ & $x$-$z$ & $x$-$y$, $x$-$z$\\
$N_{\rm cell}$ & 960 & 3840 & 3840 & 3840 & 1920\\
$N_{\rm stationary}$ &3840 & 15360 & 15360 & 15360 & 7680 \\
$N_{\rm ejecting}$ &3000 & 12000 & 12000 & 12000 & 6000 \\
$N_{\rm orbit}$ &6840 & 27360 & 27360 & 27360 & 13680 \\
\hline
\end{tabular}
\label{mass}
\end{center}
\end{table*}
\citet{Schwarzschild1979,Schwarzschild1982} proposed the orbit-superposition method to reproduce the density distribution of galaxies and build triaxial galaxy models.
The basic idea is to compute a large library of orbits in a given potential, and determine the superposition of orbits that provides the best fit to the observational
density distribution or the underlying density model. Let $N_{\rm orbit}$ be the number of the orbits in the library ($j \in N_{\rm orbit}$) and $N_{\rm cell}$ the
total number of grid cells segmenting space ($i \in N_{\rm cell}$). Further, let $O_{ij}$ denote the fraction of time spent by the $j{\rm th}$ orbit in the $i{\rm th}$
cell. The weight and mass of the $j{\rm th}$ orbit are defined by $w_j$ and $m_i$, respectively, and they are related by the following set of linear equations:
\beq\label{linear}
\sum_{j=1}^{N_{\rm orbit}} w_jO_{ij} = m_i.
\eeq
Schwarzschild's method is widely used to build spherical, axisymmetric and triaxial models for galaxies \citep{Richstone1980, Richstone1984, Pfenniger1984, RT1984,
Zhao1996_bar, Rix_etal1997, vanderMarel_etal1998, Binney2005, Capuzzo-Dolcetta_etal2007, stability}.

The $O_{ij}$ array is obtained by computing the superpositions of the $j{\rm th}$ orbit in equal time intervals $\Delta \tau_{j}$, and counting the numbers of the
output points $\nu_{ij}$ in the $i{\rm th}$ cell. After that the elements are determined according to
\beq
O_{ij} = {\Delta \tau_j \times \nu_{ij}\over \Delta \tau_j \times \nu_{j}}= {\nu_{ij}\over \nu_j},
\eeq
where $\nu_{j}$ is the total output number of the $j{\rm th}$ orbit.
The previous analyses in \citet{triaxial} and \citet{stability} used non-equal time interval outputs, given by the variation of the gravitational field strength,
$\Delta \tau_j' \sim 1/|\vec\nabla\cdot\vec{g}|^{1/2}$. In this case, the real time intervals $\Delta \tau_j'$ of the $j{\rm th}$ orbit are not constant anymore.
However, in these previous studies the unevenness of the time intervals between the outputed points along an orbit was neglected, and the $O_{ij}$ were calculated
using $O_{ij}\approx {\nu_{ij}/\nu_j}$, which systematically increases the number of output points in cuspy centres. Our present analysis does not make this
approximation. Each orbit is integrated for $100$ times its circular orbital time, $100~T_{\rm cir}$, hence the equal time interval $O_{ij}$ does not give rise
to additional inaccuracy. We will discuss the time integration of the orbits in Sec. \ref{sec-tcir}. Further details related to grid segmentation,
initial conditions, and orbital classification are separately given in Appendix \ref{OrbIntegration}.

\subsection{Integrating the orbits}\label{sec-tcir}
\begin{figure}{}
\begin{center}
\resizebox{8.5cm}{!}{\includegraphics{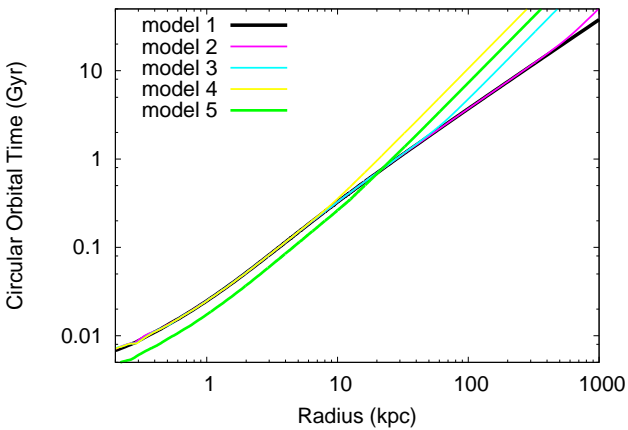}}\vskip 0.5cm
\makeatletter\def\@captype{figure}\makeatother
\caption{The circular orbital time in units of Gyr plotted against radius in units of $\kpc$, using a logarithmic scaling. The black, magenta, cyan, yellow and
green lines represent models $1$--$5$ listed in Table~\ref{mass}, respectively.}
\label{tcir}
\end{center}
\end{figure}

As stated in the previous section, all orbits are integrated for $100~T_{\rm cir}$. In Fig.~\ref{tcir}, we plot the circular orbital time against the radius for
all five models, using a logarithmic scaling. The circular orbital times of models $1$ and $4$ start to differ from each other at about $10~\kpc$. While the slope
of model $1$ is approaching unity at large radii, those of models $2$--$5$ approach a value of $3/2$ at infinity. For the isolated MOND model $1$, the circular
velocity turns constant at large radii, $v_{\rm c}=(GMa_0)^{1/4}$, hence $T_{\rm cir} \propto r$. In the case of strong external fields, however, the interpolating
function $\mu$ becomes a constant far away from the system, leading to a Newton-like behavior $T_{\rm cir} \propto r^{3/2}$.

\subsection{Smoothed solutions}
Once the orbit libraries are built and the $O_{ij}$ array is recorded, we can calculate the orbital weights in Eq.~\ref{linear}. The right-hand side of
Eq.~\ref{linear} denotes the mass of the $i{\rm th}$ cell which we obtain from Monte-Carlo simulations using the analytic density distribution. The linear
system is then solved by applying a non-negative least-squares (NNLS) method which minimises the following quantity:
\beq\label{nnls}
\chi^2 = {1\over N_{\rm cell}} \sum_{i=1}^{N_{\rm cell}}\left(m_i-\sum_{j=1}^{N_{\rm orbit}} w_jO_{ij}\right)^2.
\eeq
Furthermore, we introduce the self-consistency parameter $\delta$ \citep{Merritt_Fridman1996} as
\beq\label{selfcon}
\delta=\sqrt{\chi^2}/\overline{m},
\eeq
where $\overline{m}$ is the mean Monte-Carlo mass in cells. For self-consistent models, the value of $\delta$ is expected to quickly decrease with an
increasing number of orbits, and should be very close to zero if a large number of orbits is adopted. In Table~\ref{delta}, we list the self-consistency
parameters of all five models. We find that the isolated model and the triaxial one in a strong external field are the most self-consistent. As a result
of the broken symmetries, models in external fields should generally exhibit a lower level of self-consistency. For the axisymmetric models $2$--$4$, the
$\delta$-values are on the order of $10^{-2}$. Compared to model $5$, these systems feature only a single symmetry axis of the potential. The mass distribution
reconstructed from the orbits in the distorted potential becomes lopsided in the outer parts and does not accurately reproduce the analytic density profile.
This is in accordance with the observation that the estimated $\delta$-values grow with increasing external field strength.

\begin{table*}
\begin{center}\vskip 0.0cm
\caption{Model self-consistency and equilibrium: Here $\delta_{\rm smooth}$ and $\delta$ are the self-consistency parameters with and without smoothing,
respectively, $N_{\rm > 0}$ and $N_{\rm >0,~smooth}$ are the corresponding orbit numbers for non-zero weights (obtained from Eqs.~\ref{nnls} and \ref{smoothnnls}),
$v_{\rm rms}$ is the root-mean-square velocity in units of $\kms$, and $-2K/W$ is the virial ratio of the initial conditions. The radial-to-tangential anisotropic
ratio $\xi$ characterizes the radial instability. $L_{\rm c}$ denotes the unit-mass angular momentum with circular velocity $v_{\rm c}$ at radius $r_{\rm c}=1\kpc$,
where $L_{\rm c}=r_{\rm c}v_{\rm c}$ is expressed in units of $\kms\kpc$.
}
\begin{tabular}{lllllc}
\\
\hline
Model &1 & 2 & 3 & 4 & 5\\
\hline
$N_{\rm orbit}$ & 6840 & 27360 & 27360 & 27360  &13680 \\
$\delta$ & $1.5\times 10^{-15}$ &$1.37178 \times10^{-2}$ & $2.44238\times 10^{-2}$& $4.41140\times 10^{-2}$ & $6.2\times 10^{-5}$\\
$\delta_{\rm smooth}$ & $6.7\times10^{-5}$ & $1.37180\times 10^{-2}$ & $2.44239\times 10^{-2}$ & $4.41141\times 10^{-2}$& $6.3 \times 10^{-5}$\\
$N_{\rm >0}$       &  960 & 2378 & 2102 & 1677 & 1905 \\
$N_{\rm >0,~smooth}$ & 5995 & 2379 & 2103 & 1677 & 1949 \\
$v_{\rm rms} \left\lbrack\kms\right\rbrack$ & 225.55 & 225.20 & 224.67 & 220.28 &309.58 \\
$-2K/W$ & 1.01 & 1.02 &1.01  &1.00 &1.01 \\
$\xi$   & 1.26 & 1.74 & 1.57 & 1.65 & 1.60 \\
$L_{\rm c} \left\lbrack\kms\kpc\right\rbrack$ & 253.25 & 251.59 & 252.16 & 253.99 & 352.10\\
\hline
\end{tabular}
\label{delta}
\end{center}
\end{table*}

To construct sufficiently self-consistent models, one needs to use a large number of orbits, often far more than the number of cells $N_{\rm cell}$. The
best solutions are typically \emph{non-unique} with a very noisy phase-space distribution characterised by $N_{\rm orbit} - N_{\rm cell}$ and $N_{\rm cell}$
zero-weight and non-zero-weight orbits, respectively \citep{Merritt_Fridman1996, Zhao1996_bar, Rix_etal1997}.
As the mass distributions given by Eq.~\ref{den} are smooth, however, one would desire to select orbits in a less noisy way, i.e. with little oscillations
in the weights of neighbouring orbits in phase space. Introducing a regularisation mechanism allows one to construct a physically more plausible model. For
instance, \cite{Zhao1996_bar} smoothed the orbits by averaging the weights of the nearest $26$ neighbouring orbits when solving the NNLS.

Here we apply a simpler method of regularisation: We minimise the scatter of orbital weights by introducing a smoothing parameter $\lambda$, where
$\lambda=N_{\rm orbit}^{-2}$ is chosen as in \citet{Zhao1996_bar}. The regularisation method used here is very similar to that of \citet{Merritt_Fridman1996},
and ensures the least number of orbits with zero weights. Hence, the fluctuations of weights become smaller and the contribution of orbits to the mass
distribution becomes smoother. To this end, Eq.~\ref{nnls} is modified as
\beq\label{smoothnnls}
\chi^{2}_{\rm smooth} = \frac{1}{N_{\rm cell}}\sum_{i=1}^{N_{\rm cell}}\left(m_i-\sum_{j=1}^{N_{\rm orbit}} w_jO_{ij}\right)^2+\lambda \sum_{j=1}^{N_{\rm orbit}} w_j^2.
\eeq
Due to the regularisation, the models acquire larger $\chi^2$-values, and therefore, the solution loses part of the self-consistency. The third line of
Table~\ref{delta} shows the self-consistency parameters, $\delta_{\rm smooth}=\sqrt{\chi^2_{\rm smooth}}/\overline{m} $ (where $\overline{m}$ is the mean
Monte-Carlo mass in cells), for the smoothed models. Comparing with second line, we find that the regularisation leads to an increase of $\chi^2$ on the
order of $10^{-7}-10^{-5}$. Since the models $2$--$5$ feature symmetrised orbits, i.e., additional orbits starting from the other three octants
(the second, fifth, and sixth octants) with the same initial conditions (see Appendix \ref{OrbIntegration}), the increment of orbits is equivalent to
smoothing the orbital structure. The regularisation in Eq. \ref{smoothnnls} only slightly changes the number of orbits with non-zero weights (see Table
\ref{delta}), and thus does not considerably change the accumulative fraction of individual orbit families.

\begin{figure*}{}
\begin{center}
\resizebox{14.0cm}{!}{\includegraphics{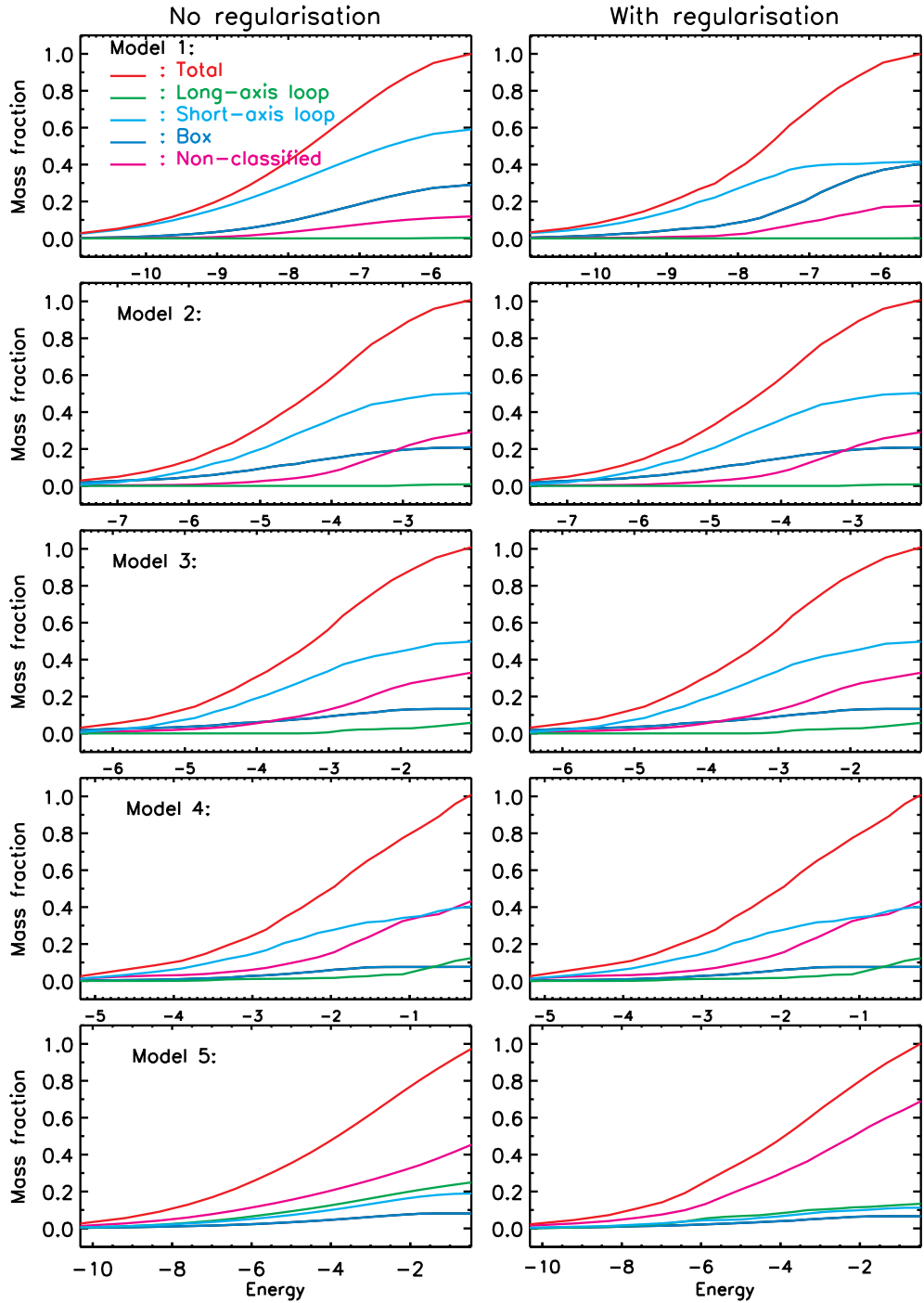}}\vskip 0.cm
\makeatletter\def\@captype{figure}\makeatother
\caption{\textbf{Left Panels:} The integrated contributions of different orbit families (for energies $< E$) to the mass as a function of energy for models $1$-$-5$
(from top to bottom panels), assuming Eq.~\ref{nnls}, i.e. no smoothing. \textbf{Right Panels:} The same as the left panels, but now assuming the regularisation given
by Eq.~\ref{smoothnnls}.}
\label{orb_family}
\end{center}
\end{figure*}

Fig. ~\ref{orb_family} shows the integrated contributions of orbits (for energies $<E$) to the system's mass with (right panels) and
without (left panels) regularisation. The individual contributions of long-axis loop (green), short-axis loop (bright blue), box (dark blue lines), and
non-classified (purple lines) orbits are plotted against the energy $E$. The fraction of box orbits in model $1$ is clearly increases when the regularisation
is applied. Such a large amount of box orbits might result in radial instability of the model. For models $2$--$5$, the smoothing procedure does not change
the fractions of orbital families noticeably; the numbers of orbits are large, and thus the solutions are smooth enough before applying the regularisation.
Reducing the number of cells and orbits to that of model $1$, we further recomputed orbits for model $2$ using Eq. \ref{smoothnnls}. The
resulting orbital structure at low energy turns out very similar to the non-smoothed solution of model $1$, indicating that the use of reflecting (symmetry)
planes changes the orbital structure for models $2$--$5$. The total amount of orbits with low angular momentum, i.e. box and non-classified orbits, are quite
large for these models. These orbits might introduce instability which will be studied in the later sections.

For the models $1$--$4$, we find that short-axis loop orbits provide a large mass fraction, comprising over $40\%$ of the total mass, even
after the regularisation. Long-axis loop orbits typically appear if an external field is applied (models $2$--$4$). The stronger the external field, the more
long-axis loop orbits emerge. Since the potential symmetry is broken at smaller radii (see Fig. \ref{isopot}), this simply reflects that stronger external fields
impact a larger fraction of orbits. The fraction of non-classified orbits follows a similar behaviour. While their fraction at fixed energies increases, the number
of box orbits is simultaneously reduced. This might imply that the external field destroys the well-defined box orbits and make them appear stochastic in phase
space.

For the triaxial model $5$, however, the orbital contributions significantly change. The smoothed result shows that the model is dominated by non-classified orbits,
making up almost 70\% of the total mass. In Newtonian gravity, \cite{Merritt_Fridman1996} have demonstrated that a galaxy constructed completely by regular orbits
is not self-consistent, and most orbits in their galaxy models are irregular, especially for the cases with strong cusps. Our result is consistent with Newtonian
models of \citet{Merritt_Fridman1996}. The two families of loop orbits are the least important components, with mass fractions close to zero at all energy levels.
Considering their total fraction in all five models, we conclude that non-classified orbits become important for systems with lower symmetry.

As solutions obtained from Schwarzschild's method are not unique, the orbital structure could change when adopting different regularisation methods. The general
conclusions, however, should remain the same.

\section{Instability of cluster galaxies}\label{nbody}
It is unknown whether quasi-equilibrium models constructed with Schwarzschild's approach are stable. The direct way to test the stability and evolution is to use
$N$-body tools. Due to the external field, the potentials of axisymmetric density profiles are lopsided, and orbits running in these potentials also become lopsided.
For an arbitrary orbit integrated in a given potential for a long enough time, the mass reproduced by this orbit will also be lopsided. Thus the uncertainty on the
model's stability increases in this case. It is an important issue to investigate the stability of MOND models in external fields since there are many elliptical
galaxies observed in clusters. In what follows, we want to take a first step into this direction by performing a kinematic analysis of the previously introduced
models, starting with $N$-body initial conditions (ICs) given by Schwarzschild's approach.

\subsection{Initial conditions and Numerical setup}
\subsubsection{Generating ICs from orbital libraries}\label{ICscheme}
We follow \citet{Zhao1996_bar} and \citet{stability} to generate the ICs for $N$-body simulations. In brief, for an $N$-particle system, the number of
particles on the $j{\rm th}$ orbit is $n_j=w_j N$, where $w_j$ is the weight of the $j{\rm th}$ orbit. Particles are placed on the $j{\rm th}$ orbit on
equidistant times given by the interval $\Delta t_j = T_j/n_j$, where $T_j$ is the total integrated time for the $j{\rm th}$ orbit (see Fig. \ref{tcir}
).\footnote{One can also randomly sample particles on the $j_{th}$ orbit from a uniform distribution. Since most of the orbits have small positive weights
in our simulations, the number of particles on $j_{th}$ are quite small. A random sampling might introduce numerical noise $\propto 1/\sqrt{n_j}$, and
could, therefore, have problems to reflect the real shape of the orbit if the weight is small. To avoid such problems, we choose an isochronous sampling.}

Our galaxy models exhibit special symmetries which significantly decrease the amount of computing work (see Table \ref{mass} and Appendix \ref{OrbIntegration}).
Also, as the NNLS selects hundreds of non-classified orbits which have not completely relaxed within $100$ circular orbital times, the systems' phase-space
symmetries are broken when placing particles on these orbits. Therefore we need to consider additional mirror particles in phase space, where the ``mirrors''
are the corresponding reflecting planes specified in Table~\ref{mass}. In the simulations, we use $10^6$ particles for each model after taking into account
these symmetry considerations. These particles represent the inner 20 mass sectors (21 mass sectors in total) of a Hernquist model. The details of segmentation of the models are shown in Appendix \ref{OrbIntegration} and in \cite{stability}.

If the ICs generated by Schwarzschild's technique are in quasi-equilibrium, the scalar virial theorem should be approximately valid, i.e. $W+2K=0$, where
\beq
W=\int\rho\vec{x}\cdot\vec{\nabla}\Phi{\rm d}^{3}x
\eeq
is the Clausius integral and $K$ is the kinetic energy of the system (Binney \& Tremaine 1987). The virial ratios $-2K/W$ and the root-mean-square velocities
of the ICs are listed in Table~\ref{delta}. We find that all five models satisfy the scalar virial theorem very well, with $-2K/W=1\pm0.02$. The $v_{\rm rms}$-values
in models $2$--$4$ are slightly smaller than in model $1$ due to the presence of external fields. Since the potential is shallower in an external field, $v_{\rm rms}
\approx\sigma$ for pressure-supported systems becomes smaller. For the strongest external field (model $4$), the $v_{\rm rms}$-value is smallest, $\approx 5~\kms$
less than in model $1$. Even in this case, the decrease of $v_{\rm rms}$ is only $2\%$, implying that the dynamics is dominated by its self-gravity for the most part. 

Consider a sizable low-mass galaxy dominated by an external field. Compared to the case of an isolated MOND galaxy or a CDM-dominated dwarf galaxy, $v_{\rm rms}$ is
expected to be much smaller. Crater II in the Local Group is such a galaxy and has recently been studied by \citet{McGaugh2016}. We know that Crater II is a very
diffuse dwarf galaxy that is dominated by the weak external field of the Milky Way at a Galactic distance of $120~\kpc$. The predicted value of $\sigma$ in this
galaxy is only $2.1~\kms$ if one accounts for this external field, but approximately twice as large for an isolated model. Given the magnitude of this effect, an
analysis of such systems could be very rewarding. A detailed study of very diffuse systems that are entirely dominated by external fields is beyond the scope of this,
paper, and will be subject to a follow-up project. 

\subsubsection{Radial instability of the models}
As discussed in section \ref{schwarzschild}, large populations of box orbits are selected to fit the underlying density distribution. In addition, the non-classified
orbits are characterised by low angular momentum, and thus highly radially anisotropic. It is therefore important to examine the radial instability of the model ICs.
To this end, we consider the anisotropy parameter $\xi=\frac{2K_r}{K_t}$ which has been introduced for spherical Newtonian systems \citep{Polyachenko_Shukhman1981,
Saha1992,Bertin_etal1994,Trenti_Bertin2006}. Here $K_r$ and $K_t=K_\theta+K_\phi$ are the radial and tangential kinetic energy components, $K_r=\sum_{\rm ip=1}^{\rm N} \frac{1}{2}m_{\rm ip}\sigma_{\rm r,ip}^2$, where the index ${\rm ip}$ runs over all particles, and $K_\theta$ and $K_\phi$ can be defined in the same manner. Spherical
Newtonian systems are unstable if $\xi$ lies above a critical value, $\xi_{\rm crit}$. Various studies have found different values of $\xi_{\rm crit}$ (e.g., \citealt[][]{May_Binney1986},
$\xi_{\rm crit} \approx 2.2$; \citealt[][]{Saha1991}, $\xi_{\rm crit} \approx 1.4$; \citealt[][]{Saha1992}, $\xi_{\rm crit} \approx 2.3$; \citealt[][]{Bertin_Stiavelli1989},
$\xi_{\rm crit} \approx 1.9$; \citealt[][]{Bertin_etal1994}, $\xi_{\rm crit} \approx 1.6$). Spherical models are generally unstable when $\xi\gtrsim 2.3$, but the models
may also be unstable for smaller values of $\xi$ if their distribution functions increase rapidly with low angular momentum. The radial instability transforms an originally
spherical system into a triaxial one, and also alters the spatial distribution of the velocity dispersion. The $\sigma(r)$-profiles become more isotropic in the centre and
radially anisotropic in the outer regions \citep{Barnes_etal2005,Bellovary_etal2008}. Moreover, axisymmetric models are unstable within a major part of the parameter space
\citep{Levison_etal1990}. For triaxial models, a collective radial instability has been studied by \citet{Antonini_etal2008}. Such instabilities are caused by the box-like
orbits, rather than non-classified ones, nor by any deviations in the model self-consistency. The radial instability causes triaxial models to become more prolate \citep{Antonini_etal2008,
Antonini_etal2009}. 

In the context of MOND, the parameter $\xi$ has been studied for spherical Osipkov-Merritt radially anisotropic $\gamma$ models (with $\gamma=0$ and $1$, where the latter
recovers the Hernquist model) \citep{Nipoti_etal2011}. It was found that a MOND system with radial anisotropy is more stable than a pure Newtonian model with exactly the
same density distribution. As the anisotropic radius in MOND is larger than that in a pure Newtonian model, a larger fraction of radial orbits can exist in the outer region
of a MOND system. The inferred $\xi_{\rm crit}$-values for MOND systems appear within the range $[2.3, ~2.6]$. Considering our galaxy models, we estimate $1.2< \xi < 1.8$
which is well below the corresponding values of $\xi_{\rm crit}$. Since the external field breaks the potential symmetries associated with these models, however, one cannot
make a conclusive statement about the presence of radial instabilities in these cases. While a stability study for isolated triaxial MOND models has been carried out by
\citet{stability}, an analysis of MOND models in external fields is still missing. In the following sections, we will present a first investigation on the stability of such
models.

\subsubsection{Numerical setup and the virial ratio}
\begin{figure}{}
\begin{center}
\resizebox{8.5cm}{!}{\includegraphics{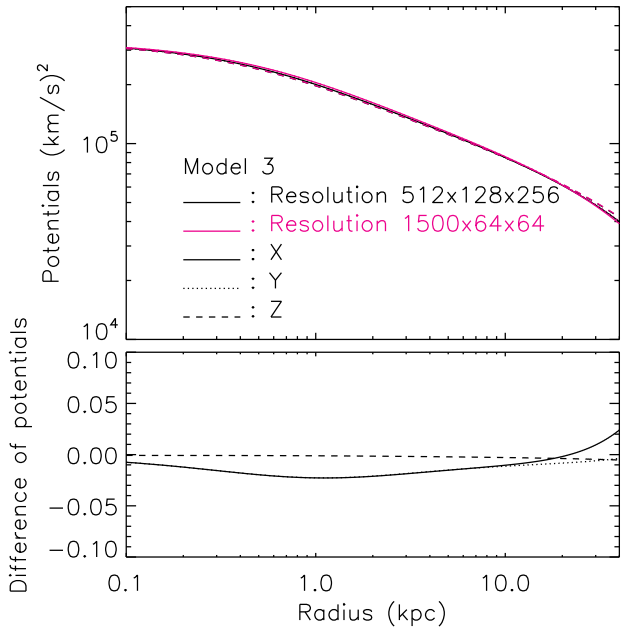}}
\makeatletter\def\@captype{figure}\makeatother
\caption{Upper panel: The static potential of model $3$ along the $x$ (solid curves), $y$ (dotted) and $z$ (dashed) axes for the Schwarzschild modeling
($\Phi_{\rm Schwarzschild}$, black curves) and for the N-body simulations ($\Phi_{\rm N-body}$, magenta curves). Lower panel: Relative deviation between
the differently resolved potentials, $\Delta \Phi = (\Phi_{\rm Schwarzschild} - \Phi_{\rm N-body})/\Phi_{\rm Schwarzschild}$.
}
\label{pot}
\end{center}
\end{figure}

Since the inclusion of external fields can be achieved by means of suitable boundary conditions, the NMODY Poisson solver does not need to be substantially altered, and
can be easily adapted to our purposes. For the simulations, we use a grid resolution of $n_r\times n_{\theta}\times n_{\phi}=1500\times64\times64$ in spherical coordinates
($r, \theta, \phi$), where the radial grid segments are defined as
\beq\label{radialbin}
r_{i} = 8\tan\left\lbrack (i+0.5)0.5\pi/(n_r+1)\right\rbrack\kpc,
\eeq
and the other two remaining grid segments are the same as in \S~\ref{model}. To reduce the computational workload, the chosen angular resolution is
lower than that adopted for the Schwarzschild modeling. The upper panel of Fig. \ref{pot} shows the static potential of model $3$ along the $x,~y$ and $z$ axes for the
high (black curves) and low resolutions (magenta curves). The two potentials agree well, with a relative difference of less than $3\%$ within $40~\kpc$ (where over
$95\%$ of the total mass is enclosed). Therefore, we do not expect any significant relaxation or evolution due to the reduction of angular resolution in the $N$-body
simulations. We also note that the potential is very similar along the $x,~y$ and $z$ axes. This indicates that MOND potentials turn out rounder than their underlying
density distributions in the regions where the dynamics is not dominated by the external field, an effect which has previously been studied in \citet{LMC}. The time unit used in the NMODY code is given by \citep{stability}
\beq
\begin{split}
T_{\rm simu} &= \left (\frac{GM}{a^3}\right )^{-1/2}\\
&= 4.7\times 10^{6}{\rm yr}\left(\frac{M}{10^{10}\Msun}\right )^{-1/2}\left (\frac{a}{1{\rm kpc}}\right )^{3/2}.
\end{split}
\label{time}
\eeq
Here the quantity $2\pi T_{\rm simu}$ has the physical meaning of one Keplerian time at a radius of $r=a$, and the system's total mass $M$ is expressed in units of
$1.0\times 10^{10} \Msun$.

It is well known that the typical size of galaxy clusters is on the scale of several Mpc. However, their central regions where there exist strong and nearly
uniform gravitational backgrounds are much smaller. To give a rough estimate, the size is typically one order of magnitude smaller than the size of the cluster
which is around $0.1$ Mpc. Galaxies are accelerated in an almost constant field at this scale. Converted into a physical time scale where this approximation
holds, this gives around $60~T_{\rm simu} \sim 0.3$ Gyrs. Of course, real galaxies are accelerated within inhomogeneous fields, but this general case is still
too complex to be modeled at present, and most of the physics we are interested in at the moment can explored in a constant background. More details about the
time steps used in the code are discussed in \cite{stability}. 
We have simulated our models up to $120~T_{\rm simu}$ (twice the value of $60~T_{\rm simu}$) to examine the systems' behaviour beyond the actual simulation time
interval. In what follows, we will restrict the discussion to within $60~T_{\rm simu}$; only virial ratios are presented for the fully simulated range of $120~T_{\rm simu}$.

\begin{figure}{}
\begin{center}
\resizebox{7.5cm}{!}{\includegraphics{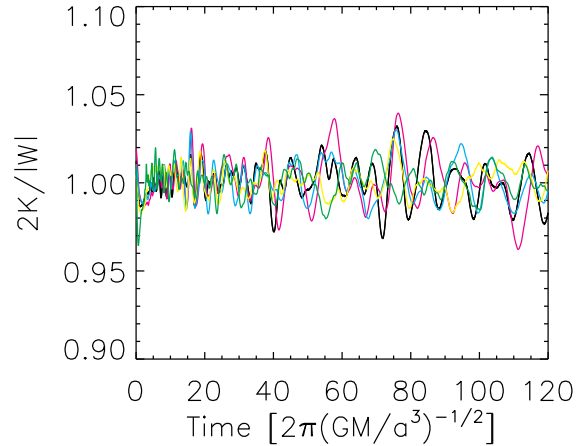}}
\makeatletter\def\@captype{figure}\makeatother
\caption{Evolution of the virial ratio: The black, magenta, cyan, yellow and green lines represent models $1$--$5$, respectively.
Here $2\pi \left (GM/a^{3}\right )^{-1/2}$ is the Keplerian rotation time at the length scale $a=1\kpc$ for a total mass $M$.}
\label{virial}
\end{center}
\end{figure}

As mentioned above, the external field models are not exactly self-consistent with respect to the original analytic density profile. The virial
ratios of the ICs slightly deviate from unity ($\pm 2\%$), and we expect these quasi-equilibrium ICs to quickly relax to dynamically virialsed systems at the beginning
of the simulations. Figure~\ref{virial} shows the evolution of the virial ratio $2K/|W|$ within $120~T_{\rm simu}$. As can be seen from the figure,
the models revirialise within a few simulations times and then the virial ratios oscillate around $1$ for all five models, with an oscillation amplitude of roughly
$0.05$ within $120~T_{\rm simu}$. Since the overall residuals of virial ratios at $T=0$ are at the level of few percent, the deviation from the
exact equilibrium state is an minor effect. Hence we conclude that the virial theorem is valid for all considered models.

\subsection{Mass distribution}
\label{massdist}
The presence of an external field gives rise to a lopsided potential. Thus the mass density will redistribute inside the total potential until
the density with its associated (internal) potential reaches an equilibrium configuration. In the following, we want to address how far-reaching
this evolution in an external field is.

\begin{figure}{}
\begin{center}
\resizebox{8.6cm}{!}{\includegraphics{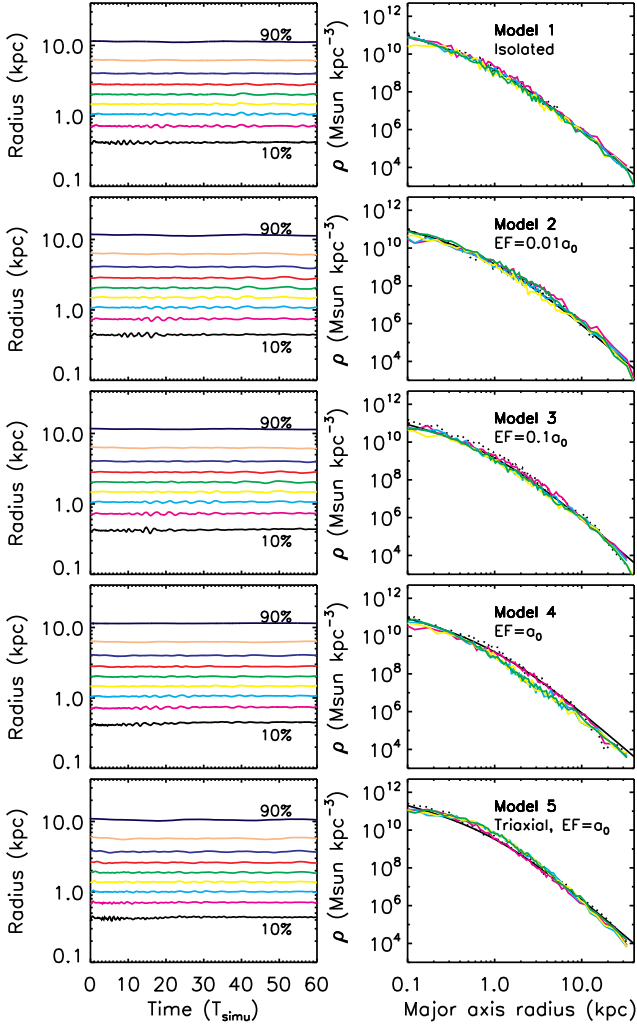}}\vskip 0.2cm
\makeatletter\def\@captype{figure}\makeatother
\caption{\textbf{Left panels}: The evolution of Lagrangian radii for models $1$--$5$ (from top to bottom, respectively). \textbf{Right panels}:
Density distribution along the galaxy major axes. The analytic initial density and the density constructed from $N$ particles correspond to the
black solid and dashed lines, respectively. The violet, blue, yellow and green lines correspond to simulation times of $15$, $30$, $45$ and
$60~T_{\rm simu}$, respectively.}\label{enclosed}
\end{center}
\end{figure}

The \textbf{left panels} of Fig. \ref{enclosed} illustrate the spherical radii $r_0(M)$ enclosing different fractions of the total mass $M$, i.e., Lagrangian radii, increasing from 10\% to 90\% in steps of 10\%. For models $1$--$3$, these radii
are very stable, showing only tiny oscillations within $60$ simulation times. Only the innermost $10\%$ mass radii of models $4$ and $5$ slightly
increase by about 15\%. This implies that the global radial mass distributions do not significantly evolve with time. For
models in strong external fields, the radial mass distributions slightly decrease in the innermost regions and stay almost constant elsewhere. 

Note that there are less symmetries for models $2$--$5$, and thus there are less mirror particles in phase space. Such a situation could result
in a self-rotation which cannot be canceled due to the lack of counteracting mirror particles. This is especially true for the outer regions where
the impact of the external field starts to become important. In \S\ref{section44}, we will further comment this issue. As a consequence, the major
axes of models $2$--$5$ may not coincide with the $x$-axis anymore.

To this end, we determine the system's principal axes according to the following approach. Starting from an initial guess $p=q=1$, we consider all
particles within an ellipsoid defined by $x^2+(y/p)^2+(z/q)^2=r^2$. These particles are then used to compute the components of the weighted moments
of inertia tensor given by
\bey\label{mit}
\ixxt &=&\frac{\sum_i m_i(y_i^2+z_i^2)/r_i^2}{\sum_i m_i},\nonumber\\
\ixyt &=&\frac{\sum_i -m_ix_i y_i/r_i^2}{\sum_i m_i},
\eey
and similar expressions for the other components. Here we adopt the weighted moments of inertia tensor to mitigate the noisy contribution of particles
at larger radii. The resulting inertia tensor is diagonalised, yielding eigenvalues $\ixxtnew,~\iyytnew$ and $\izztnew$, where the primed coordinate
system refers to the corresponding eigenframe. The associated principle axes,
\bey\label{abc}
\at(r)&=&\sqrt{[\iyytnew +\izztnew -\ixxtnew ]/2},\nonumber\\
\bt(r)&=&\sqrt{[\ixxtnew +\izztnew -\iyytnew ]/2},\nonumber\\
\ct(r)&=&\sqrt{[\ixxtnew +\iyytnew -\izztnew ]/2},
\eey
are used to determine new values of $p\equiv \bt /\at$ and $q\equiv \ct /\at$, and the particle coordinates are rotated into the inertia tensor's eigenframe.
The procedure is repeated iteratively until both axial ratios $p$ and $q$ converge to a relative deviation of less than $10^{-3}$. 

The \textbf{right panels} of Fig.~\ref{enclosed}, show the models' density distributions along their major axes obtained from the principal axes at a
spheroidal radius enclosing $90\%$ of the total mass, i.e. $r_{90\%}$.\footnote{The oscillations of the radial densities emerge from numerical noise.
The radial densities along the major axes are computed from the $N$-body simulation grids which are closest to the major axes. Approximately, there
are $N_P/(n_\theta \times n_\phi) \approx 250$ particles along the major axes, and Eq. \ref{radialbin} is used for radial binning. The $m_{th}$ radial
bin contains $n_{p,m}$ particles, and the associated numerical noise is $\approx 1/\sqrt{n_{p,m}}$.} The initial density distributions (black dotted
curves) along major axes agree well with the analytical densities of Hernquist profiles (black solid curves). The density profile of model $1$ does not
change substantially during the simulation. For models embedded in external fields, the density profiles quickly evolve to new profiles within $15-30~T_{\rm simu}$
and reach stable states afterwards. For model $2$, the density along major axes within $1$--$5~\kpc$ increases about $40\%$ relative
to the initial configuration. For models $3$ and $4$, the densities along the major axes decrease over the range $r\in [1.0,~10.0] ~\kpc$ by approximately
$50\%$ and $60\%$, respectively. The density evolution becomes more pronounced for stronger external fields (model $4$). The density profile of model $5$
increases by about $60\%$ in the inner region ($r<5.0~\kpc$), but there is no noticeable evolution in the outer parts. The differences
in the evolution of density profiles along the major axes between models $2$--$5$ are generally due to varied strengths and directions of the external field.

\subsection{Axial ratios}\label{sec-shape}
\begin{figure*}{}
\begin{center}
\resizebox{14.0cm}{!}{\includegraphics{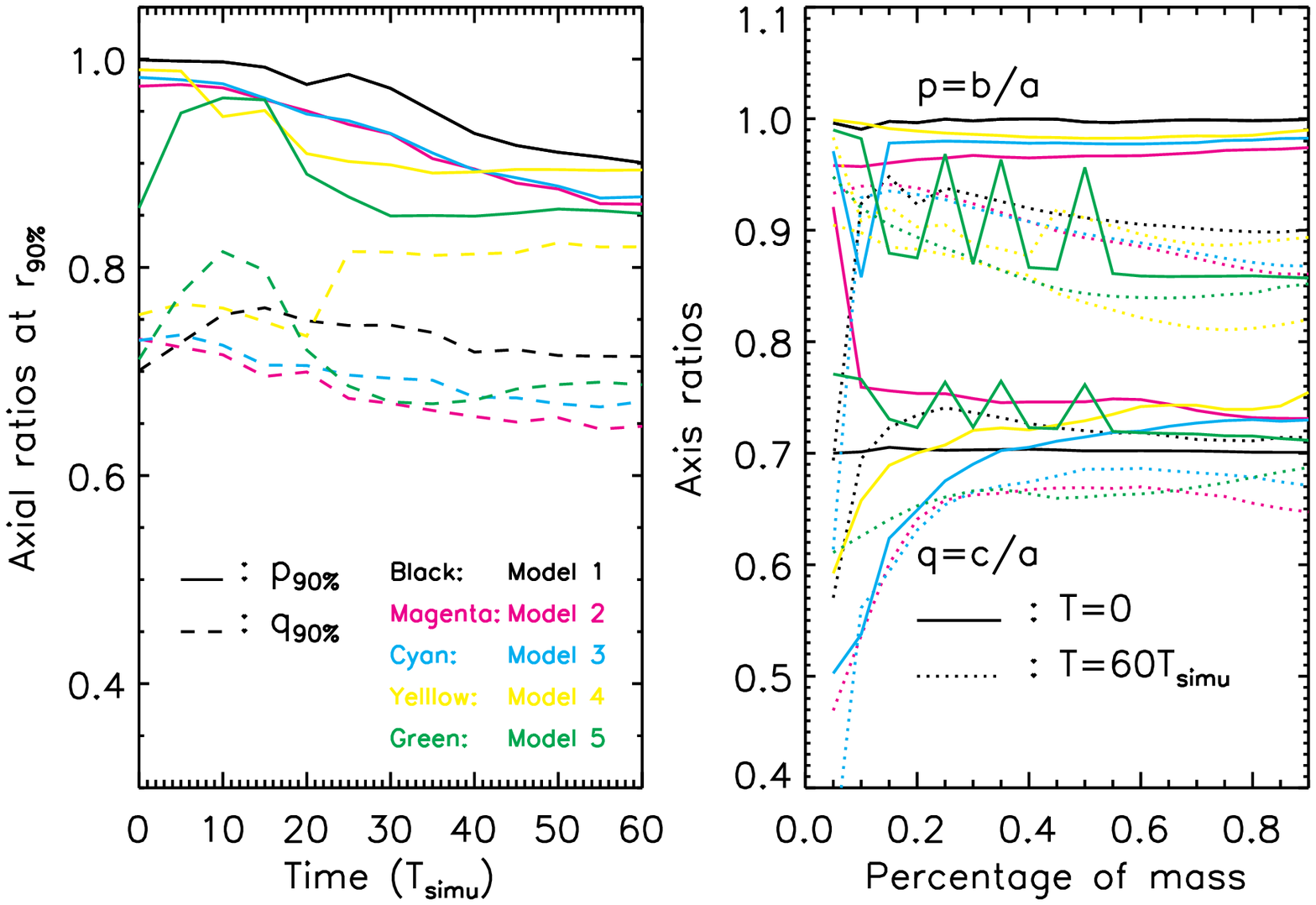}}\vskip 0.5cm
\makeatletter\def\@captype{figure}\makeatother
\caption{\textbf{Left panel}: Time evolution of axial ratios at radii enclosing $90\%$ of the total mass. The colours black, magenta, cyan, yellow and green
represent the models $1$--$5$, respectively, and illustrate the axial ratios $\pr=\btr/\atr$ (solid lines) and $\qr=\ctr/\atr$ (dashed lines). \textbf{Right panel}:
Axial ratios of the five models as a function of enclosed masses. The upper set of curves refers to $p=\bt /\at$, and the lower set to $q=\ct /\at$. The solid
and dotted curves represent initial ($T=0$) and final ($T=60~T_{\rm simu}$) ratios, respectively. The colours are the same as in the \textbf{left panel}.}
\label{principle-axes}
\end{center}
\end{figure*}

Using the iterative procedure introduced in the previous section, we are able to obtain the shapes of the systems. The \textbf{left panel} of Fig.~\ref{principle-axes}
shows the axial ratios of all five models defined for $r_{90\%}$, $\pr=\btr/\atr$ and $\qr=\ctr/\atr$. The ratios $\pr$ and $\qr$ significantly evolve for all models
during $60~T_{\rm simu}$, where the evolution of $\pr$ is generally stronger than that of $\qr$. For all models, the values of $\pr$ become considerably smaller. We
observe that the axisymmetric models evolve into a triaxial configuration, and the initially triaxial model turns slightly more prolate. 

Unlike the isolated triaxial models with axial ratios $\atr :~\btr :~\ctr~=~1.00:~0.86:~0.70$ studied in \citet{stability}, an instability appears for model $1$
(black curves) within $60~T_{\rm simu}$, yielding final axial ratios $\atr :~\btr :~\ctr~=~1.00:~0.90:~0.71$. There are oscillations of $\qr$ around $0.7$ with
an amplitude of $\pm0.05$ within $60~T_{\rm simu}$. This suggests that the triaxial model in MOND is more stable than the axisymmetric one when the system is isolated.

For the models $2$--$4$ (axisymmetric models in external fields), however, the axial ratios have initially the same value as in model $1$,
$\atr:~\btr:~\ctr ~= ~1.0:~1.0:~0.7$. With external fields pointing into the (diagonal) negative $z$-$x$- or negative $x$-$z$-direction, $\atr$ and $\btr$ start
to separate from each other at the beginning of the simulations. For the models $2$ and $3$, $\pr$ decreases gradually to $0.85$, and $\qr$
drops slowly to around $0.65$ at $60~T_{\rm simu}$. For model $4$, the values of $\pr$ decrease to around $0.9$ within the first $20~T_{\rm simu}$ and stay
constant afterwards. The ratio $\qr$ oscillates around $0.75$ within $20~T_{\rm simu}$, and then jumps to $0.82$ at $25~T_{\rm simu}$. There is no significant
evolution of $\qr$ at later times. The different evolution of axial ratios between the models $2,3$ and model $4$ is driven by the strength of the external
field which is strongest for model $4$.

The models $2$--$4$ eventually evolve into triaxial systems, and the final state is reached after $\sim 55~T_{\rm simu}$ for the models $2,3$ and $\sim 25~T_{\rm simu}$
for model $4$.
Note that the model analysis is made after rotating the systems into the reference frame defined
by their principle axes. Hence, any effects of pattern rotation (see the evolution of angular momentum in Fig. \ref{inertia}) due to the external field are eliminated.

For model 5 (green lines), the situation appears simpler. Initially, the axial ratios are approximately $\atr :~\btr :~\ctr~=~1.00:~0.86:~0.71$, which is close to
the isolated case discussed in \cite{stability}. Both $\pr$ and $\qr$ start to increase within $10~T_{\rm simu}$, but then decrease until $25~T_{\rm simu}$,
where $\atr :~\btr :~\ctr~=~1.00:~0.87:~0.69$. Again, there is no significant evolution beyond this point, and one finds $\atr :~\btr :~\ctr~=~1:00:~0.85:~0.69$ at
$60~T_{\rm simu}$. 
Since the external field for this model points into the $x$-direction, i.e. perpendicular to $y-z$ plane, the curves for the components along $y$- and $z$-axis display
a very similar running behaviour. The instability appears within the first $25~T_{\rm simu}$ for the models $4$ and $5$, after which they become stable.
We also note that the evolution for the models $1$--$3$ is not too different, since the external field effect in these cases (models $2$ and $3$) are mild. Again, this
shows that the evolution of axial ratios is closely related to the external field strength.

The axial ratios as a function of enclosed masses are presented for all five models in the \textbf{right panel} of Fig. \ref{principle-axes}. For the initial models
(solid curves), the axial ratio agrees well with the analytic density distribution and does not evolve considerably with increasing mass for the isolated model (model
$1$). For the models in external fields (models $2$--$4$), there are deviations between the analytical and model's principal axes in the inner regions. There are small
oscillations or spikes for the resulting $p=\bt /\at$ and $q=\ct /\at$ in model $5$, with amplitudes $< \pm 0.1$. These spikes are purely numerical features and emerge
from the iterative procedure, used to determine the principal axes, which is quite sensitive to the chosen initial guess in this case. 

The initial axial ratios of the models $2$--$5$ deviate from the original analytic models, especially within $r_{20\%}$, the Lagrangian radius enclosing
20\% of the total mass. These effects are caused by the offsets between the cusp centres and the center of masses (CoMs) of the models \citep{Wu_etal2010}. Considering
models with an external field applied, the systems are accelerated and moving due to the constant background field. To keep the angular resolution of the simulation at
a reasonable size, the galactic centre has to be placed at the computational grid's centre. In our simulations, we transformed the coordinates at every time step, moving
the centre of mass (CoM) to the grid centre and changing to the frame where its velocity is zero. The CoM does not need to coincide with the galactic centre, i.e. the
centre of the cusp. Due to the lopsided potential, such an offset within the density distribution can develop within MOND. However, this will happen neither in Newtonian
gravity nor for isolated models in MOND. To better understand this effect, \citet{Wu_etal2010} have designed a simple experiment: a spherically symmetric Plummer model
is placed into an homogeneous external field. In the case of Newtonian gravity, the superposition principle applies, and the whole system is equally accelerated along the
external field's direction. Hence the position of the CoM does not move in the internal system. For MOND, however, the internal gravity is determined by both the external
field and the internal matter distribution. Hence, the outer parts of the galaxy will become lopsided, generating an additional external field itself that will act on the
inner part. The additional field will cause the CoM to slightly shift away from the point where gravity equals to the external field, i.e. the point where internal gravitational
forces cancel. Besides discrepancies in the innermost regions, the axial ratios as a function of enclosed masses differ from the analytic axis-symmetric models in external fields
also at larger Lagrangian radii. For instance, $p \approx 0.96$ and $q\approx 0.74$ in model $2$. Such deviations show that our ICs for the $N$-body simulations do not
perfectly describe the shapes of the original analytic axis-symmetric models embedded in external fields. This is likely due to departures from self-consistency as the
corresponding $\delta$-parameters of the models $2$--$4$ are around a few percent and the ICs of the models are in quasi-equilibrium. The shapes of models $1$ and $5$
(again, in the outer regions) agree very well with that of the analytic models, and their self-consistency parameters are about three orders of magnitudes smaller.

At $T=60~T_{\rm simu}$ (dotted curves), the values of $p$ for the models $1$--$3$ drop to $\approx 0.94$ in the inner region and to $\approx 0.86-0.90$ at $r_{90\%}$.
The values of $q$ do not change as much as that of $p$, with amplitudes $< ^{+0.02}_{-0.05}$. This agrees with the results in the left panel of Fig. \ref{principle-axes}.
The models $1$--$3$ evolve into triaxial configurations due to the radial instability. The evolution of model $4$ is much more striking. The value of $p$ for the innermost
particles within $r_{5\%}$ is about $0.98$. It then decreases to $0.88$ at $r_{40\%}$ and jumps up to $0.92$ for an enclosed mass of $45\%$. The values of $p$ stay almost
constant in the outer region of the model. The sudden increase of $p$ is an effect of the offset between the CoM and the cusp center. The values of $q$ are larger than
that of the ICs. At $r_{5\%}$, $q\approx 0.90$ and decreases to around $0.82$ at $r_{90\%}$. The instability is mainly caused by non-classified orbits (see the orbital
fraction in the {\bf right panels} of Fig. \ref{orb_family}) and changes the shape of the model. Model $4$ turns out more prolate when compared to the models $1$--$3$.
For model $5$, the axial ratio $p$ does not evolve substantially in the strong external field. The values of $q$ are generally about $0.05$ lower than in the ICs.
Compared to model $4$, the shape evolution of model $5$ is considerably suppressed. There are two possible reasons: either a triaxial model is more stable than an
axisymmetric model in MOND, or the symmetry of model $4$ is broken in a more complex way since the external field direction is along the negative $x$-$z$ diagonal.

To summarise, our results imply that
\begin{enumerate}
\item an axisymmetric model with and without the external field effect is not stable in MOND, and that
\item the principle axes, describing the shape of a galaxy, generally evolve more substantially in the presence of a strong external field, and that
\item radial instability caused by both box orbits and non-classified orbits with low angular momentum can change the axial ratios, i.e. the shape of
a galaxy. The origin of the instability will be studied later in Sec. \ref{oriinsta}. 
\end{enumerate}

\subsection{Kinetics}
From \S \ref{massdist}, we already know that the major axes of inner regions of the galaxy models in an external field are slightly expanded compared to the isolated case. 
Additional information on these models can be obtained by exploring their properties in the full phase space. To study the dynamical evolution of our systems, we calculate the radial velocity dispersion $\sigma_r(r)$ and the anisotropy parameter
\beq
\beta(r) = 1-{\sigma_{\theta}^2 + \sigma_{\phi}^2\over 2\sigma_r^2},
\eeq
where $r$ is defined in Eq.~\ref{den} and $\sigma_{\theta}$, $\sigma_{\phi}$ are the tangential and azimuthal velocity dispersions, respectively.

\begin{figure}{}
\begin{center}
\resizebox{9.0cm}{!}{\includegraphics{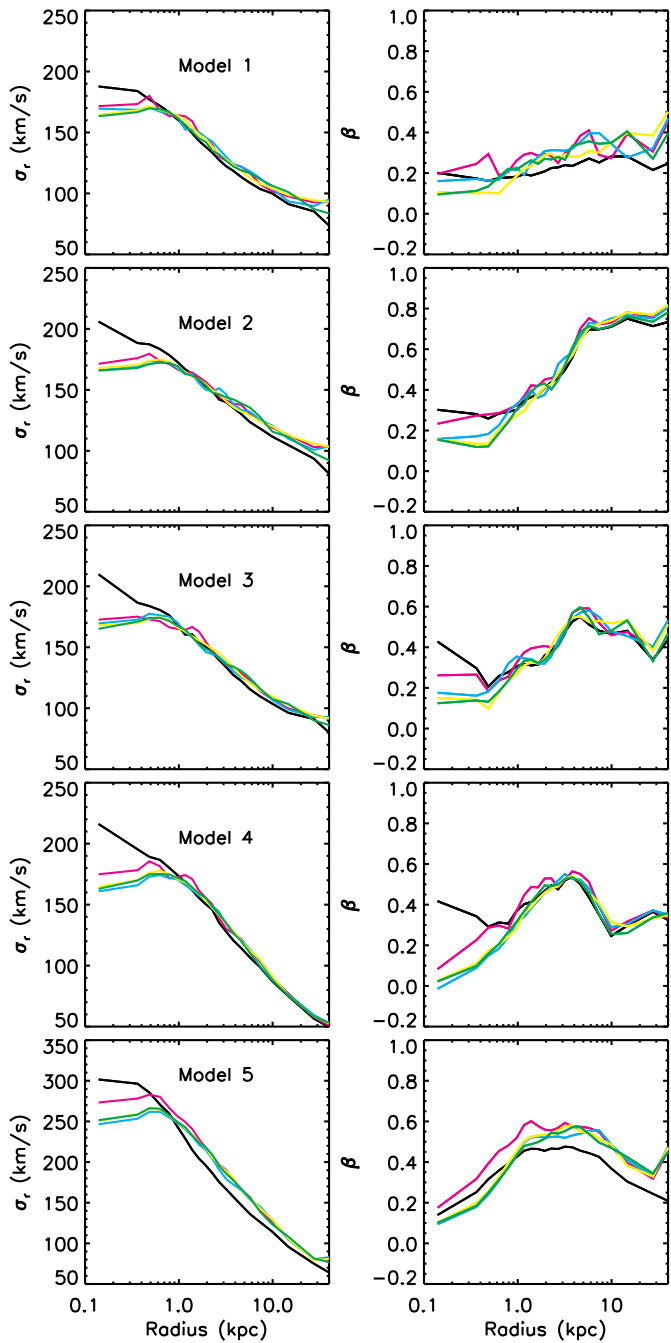}}\vskip 0.5cm
\makeatletter\def\@captype{figure}\makeatother
\caption{\textbf{Left panels}: Time evolution of the radial velocity dispersion. The panels (from top to bottom) show the results for the models $1$--$5$.
\textbf{Right panels}: The anisotropy parameter $\beta$. The black lines represent the ICs and the different colours are defined as in Fig.\ref{enclosed}.}\label{sig}
\end{center}
\end{figure}

The \textbf{left panels} of Fig.~\ref{sig} show the time evolution of the radial velocity dispersion $\sigma_r(r)$.\footnote{The radial velocity dispersions
are plotted in 21 radial bins of equal mass (see Sec. \ref{OrbIntegration}). Due to the increased particle number in these bins, the numerical noise is much
smaller than in the \textbf{right panels} of Fig. \ref{enclosed}. The same binning procedure is applied for other quantities that are discussed hereafter.}
Obviously, all models start off with an instability and seem to reach a stable state after a short evolutionary phase. The dispersions $\sigma_r(r)$ of all
models converge to stable states within $15-30 T_{\rm simu}$. In addition, we find that $\sigma_r(r)$ is flattened in the inner parts ($< 1.0$~kpc) of all
systems, with $\sigma_r(r<1.0\kpc) \approx 170\kms$ for the models 1--4. As expected, the models in external fields (i.e. models $2-5$) exhibit more evolution
than the isolated model $1$. For the same density model (models $1$--$4$), a stronger external field also leads to more evolution in the $\sigma_r(r)$-profiles. Velocity dispersions in model 5 are much larger, since the mass of model 5 is a factor of $2$ larger than other models.
The $\sigma_r(r)$-profile of the triaxial model $5$ slightly increases after the relaxation within $30~T_{\rm simu}$. At larger radii, dispersions $\sigma_r(r>1\kpc)$
are about $10-20~\kms$ higher than in the initial state. The inner region of $\sigma_r(r)$ for model $5$ decreases from $300~\kms$ to $250~\kms$ within $30T_{\rm simu}$,
and the profile also appears flat within $1~\kpc$. The evolution from the ICs to the final stable state indicates a radial instability which might originate from box
orbits or from non-classified orbits with low angular momentum.

\begin{figure}{}
\begin{center}
\resizebox{8.5cm}{!}{\includegraphics{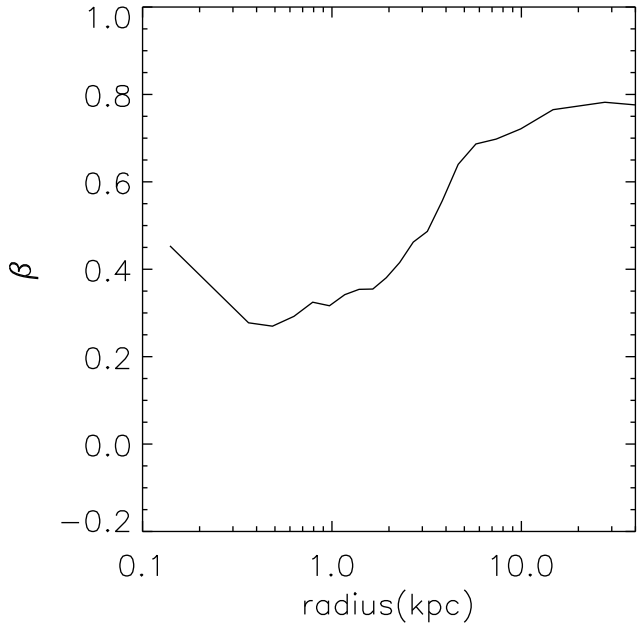}}\vskip 0.5cm
\makeatletter\def\@captype{figure}\makeatother
\caption{The anisotropy profile of model $\tilde 1$.}\label{smoothmodel1}
\end{center}
\end{figure}

For the axisymmetric models $1$--$4$, the profiles of the anisotropy parameter (right panels of Fig. \ref{sig}), $\beta(r)$, are quite different for the isolated
and non-isolated models. In the case of an isolated galaxy, model $1$, $\beta(r)$ is initially almost constant, $\approx 0.2$, in accordance with the results of
\cite{stability}. When there is an external field (models $2$--$4$), however, the initial $\beta(r)$-profiles are no longer constant. In general, $\beta(r)$ declines
in the inner region and start to increase from $\simeq 0.5~\kpc$ to $\simeq 4~\kpc$. The model with weakest external field has the highest radial anisotropy at
$4\kpc$, $\beta(r \ge 5\kpc)\approx 0.8$. The $\beta(r)$-profile stays almost constant in the outer region where $r> 5~\kpc$. Given that the
external field is very weak in model $2$, it seems unexpected that there is such a large difference between the models $1$ and $2$. The main reason is that the
numbers of cells in the grids and the numbers of orbits in the orbital libraries are different. As previously mentioned, there are three times more cells and
orbits in the models $2$--$4$ (see Table \ref{mass}). The increment of cells and orbits is equivalent to additional smoothing for the orbital structure. To
examine the impact of increasing the number of cells and orbits, we constructed a new model, labelled $\tilde 1$, by launching orbits from the same octants as
in model $2$ and increasing the number of orbits to $27360$. Generating the corresponding $N$-body ICs as before, we then computed the anisotropy profile of model
$\tilde 1$ and the result is shown in Fig. \ref{smoothmodel1}. We find that $\beta(r)$ for model $\tilde 1$ is very similar to that of model $2$, confirming that
the different numbers of cells and orbits yield the observed discrepancy between the $\beta(r)$-profiles of models $1$ and $2$ at $T=0$. 

When the external fields become stronger (models $3$ and $4$), the maximal values of $\beta(r)$ are about $0.6$ at $4~\kpc$ and the $\beta(r)$-profiles
fall off again at larger radii. The $\beta(r)$-profiles for ICs with the strongest external fields, e.g. model $4$, have the smallest values at large
radii among the external field models. As the external field amplitude increases, the radial anisotropy is reduced in the outer region of the models. 

It is important to keep in mind that the solution of Schwarzschild's method is not unique. Therefore, possible changes in the orbital structure could
lead to different anisotropy profiles of the velocity dispersion. For all the models in this work, however, the regularisation method is the same. It
seems safe to conclude that for models of the same mass derived from the regularisation in Eq. \ref{smoothnnls}, stronger external fields will generally
lead to less radially anisotropic models at larger radii.

The coloured curves show the $\beta(r)$ profiles at different simulation times. For the isolated model, the $\beta(r)$ profile does not evolve very much,
which is again similar to the results presented in \citet{stability}. For the axisymmetric models embedded in external fields along the diagonal $x$-$z$
direction, $\beta(r)$ does not substantially evolve in the outer regions where $r>1.0~\kpc$, but decreases in the inner region within $15-30~T_{\rm simu}$
after which it remains stable. The models with the strongest external field show the most prominent evolution within $1~\kpc$. At $60~T_{\rm simu}$, $\beta(r)$
is nearly zero in the central region, i.e. the centre is basically isotropic. This is consistent with the left panels of Fig. \ref{sig}. Model $5$ eventually
becomes less radially anisotropic in the inner region and more anisotropic at larger radii. Since the fraction of box orbits is quite small, the numerous
non-classified orbits might be the origin of radial instability for the triaxial model $5$. 

Our results are consistent with the shape evolution illustrated in Fig.~\ref{principle-axes}. Systems embedded in external fields appear stable after an
early evolution caused by radial instability. Nevertheless, more dynamical quantities need to be investigated before making any such statements. Further dynamical studies related to kinetic energy and angular momentum are presented in Appendix \ref{furtherdyn}.

\section{The origin of the instability}\label{oriinsta}
\subsection{Box orbits removed}\label{sec-nobox}
\begin{table}
\begin{center}\vskip 0.00cm
\caption{Parameters of the new orbital library after removing the box orbits. The number of total orbits is denoted by $N_{\rm orbit}'$, $\delta_{\rm smooth}$
is the self-consistency parameter, and $N_{\rm >0,~smooth}'$ is the number of orbits with non-zero weights. The values of the virial ratio, $-2K/W$, and of
the radial stability parameter, $\xi$, are also listed.}
\begin{tabular}{llllc}
\\
\hline
Model &$1'$ & $2'$ & $3'$& $4'$  \\
\hline
$N_{\rm orbit}'$ & 4178 & 19361 & 21424 & 23372 \\
$\delta_{\rm smooth}$ & $1.52\times10^{-2}$ & $5.33\times 10^{-2}$ & $4.32 \times 10^{-2}$ & $ 4.90 \times 10^{-2}$\\
$N_{\rm >0,~smooth}'$ & 617 & 1517 & 1601 & 1465  \\
$-2K/W$ & 1.01 & 1.01 &1.01  &1.00  \\
$\xi$   & 0.98 & 1.58 & 1.79 & 1.65 \\
\hline
\end{tabular}
\label{tab-nobox}
\end{center}
\end{table}

\begin{figure*}{}
\begin{center}
\resizebox{14.0cm}{!}{\includegraphics{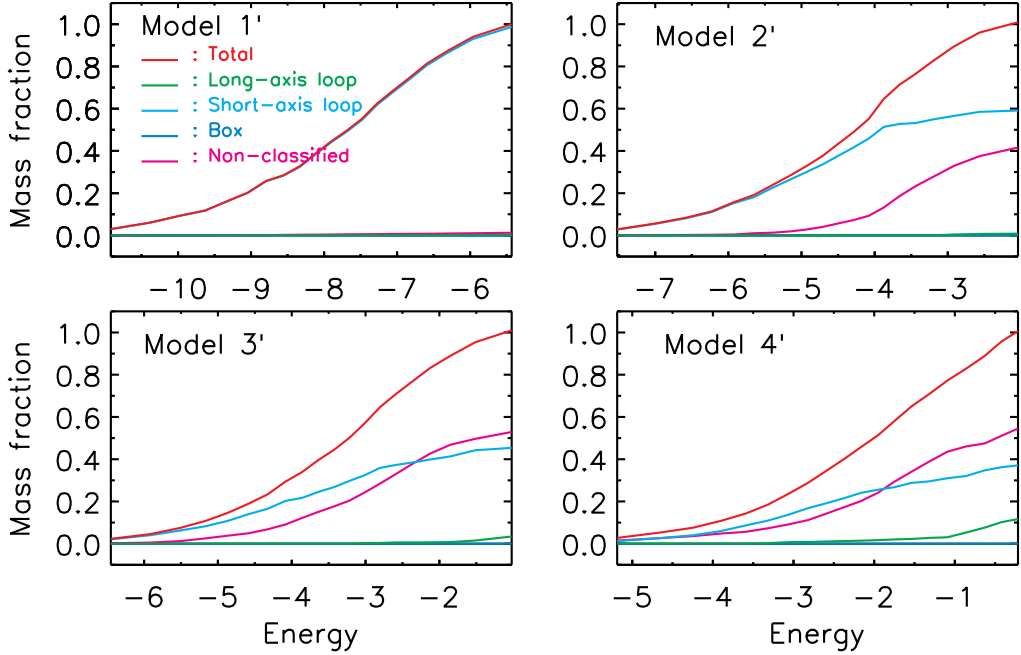}}\vskip 0.5cm
\makeatletter\def\@captype{figure}\makeatother
\caption{The orbital structure of the new models $1'$--$4'$ selected by Eq. \ref{smoothnnls} after removing box orbits from their orbital library.
}
\label{nobox_of}
\end{center}
\end{figure*}

The analysis of Sec. \ref{nbody} showed that axisymmetric models are unstable and that the models exhibit large fractions of box and non-classified
orbits with low angular momentum. In order to examine the origin of the instability, we shall perform a further test here. In our test, the box orbits
are removed from the global orbital library of the models $1$--$4$. Thus box orbits no longer contribute to the mass of these systems, and the corresponding
models are labelled through a prime, i.e. $1'$, $2'$, $3'$ and $4'$. The numbers of orbits, $N_{\rm orbit}'$, in the new libraries are listed in Table
\ref{tab-nobox}. There are at least four thousand orbits for the axisymmetric models. The best-fit solutions of Eq. ~\ref{smoothnnls} are then computed
based on the new orbital library. The new $\delta_{\rm smooth}$-parameters are on the order of $10^{-2}$ and all larger than for models which include
box orbits. This is reasonable because removing these orbits is equivalent to introducing additional constraints on the weights for the original orbital
library. The number of non-zero weights in the new models is smaller than in the original ones. Fig. \ref{nobox_of} shows the orbital structure of the
four new axisymmetric models. In these models, the fractions of short-axis loop orbits generally increases. The isolated system (model $1'$) is almost
entirely comprised of short-axis loop orbits. The fraction of non-classified and long-axis orbits increases with growing external field strength. Model
$2'$ is dominated by short-axis ($\approx 60\%$) and non-classified orbits ($\approx 40\%$). The fraction of non-classified orbits in models $3'$ and $4'$
(contributing over $50\%$ of the total mass) are larger than that of short-axis orbits. The fraction of long-axis loop orbits reaches up to $12\%$ in model $4'$. 

Generating $N$-body ICs according to the scheme presented in Sec. \ref{ICscheme}, we have conducted the same stability tests as before. The virial ratios of
the new ICs are listed in Tab. \ref{tab-nobox}. Again, all models satisfy the scalar virial theorem very well, which implies that the new ICs are close to
equilibrium. The first step of the stability test is to examine the $\xi$ values of the new models (see Tab. \ref{tab-nobox}). We find that $\xi << \xi_{\rm crit}$
for model $1'$ while $\xi \in [1.58, ~1.79]$ for the other ones. These values lie within the $\xi$-range estimated for the original models $2$--$4$. While
model $1'$ is expected to be stable, the situation for the other models is unclear and needs to be analysed in more detail.

\begin{figure}{}
\begin{center}
\resizebox{8.5cm}{!}{\includegraphics{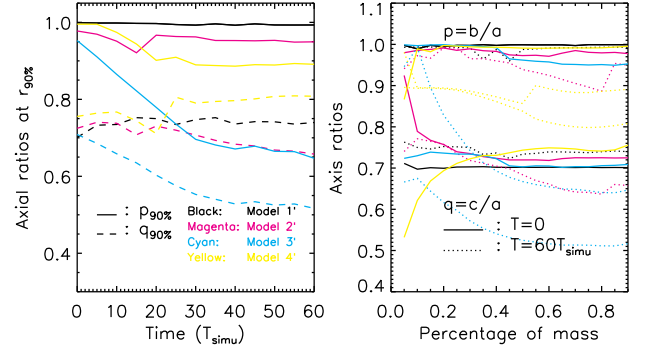}}\vskip 0.5cm
\makeatletter\def\@captype{figure}\makeatother
\caption{\textbf{Left panel}: Time evolution of axial ratios at $r_{90\%}$ for models $1'$ (black), $2'$ (magenta), $3'$ (cyan)  and $4'$ (yellow).
The solid, dotted and dashed lines denote the different components $\at$, $\bt$, and $\ct$, respectively. \textbf{Right panel}: Axial ratios of the
four models as a function of enclosed mass. The upper set of curves refers to $p=\bt /\at$ and the lower set to $q=\ct /\at$. The solid, dotted and
dashed curves represent initial ($T=0$) and final ($T=60~T_{\rm simu}$) axial ratios.
}
\label{nobox_ar}
\end{center}
\end{figure}

The shape evolution of the new models at $r_{90\%}$ for $T=0$--$60~T_{\rm simu}$ is presented in the left panel of Fig. \ref{nobox_ar}. In contrast to the
original isolated model with box orbits, the global axial ratios for model $1'$ does not evolve with time. The axial ratios as a function of enclosed mass
are shown in the right panel of Fig. \ref{nobox_ar}. The black dotted curves confirm that the shape of model $1'$ does has not changed after $60~T_{\rm simu}$.
Therefore, model $1'$, which is mostly comprised of short-axis loop orbits and has only a tiny fraction of non-classified orbits, is stable.
The global axial ratios, $\pr$ and $\qr$, of the models $2'$--$4'$ evolve similarly to that of the original models $2$--$4$, indicating that these models are
not stable. The instability might arise from the large population of non-classified orbits which contain little angular momentum. The right panel of Fig. \ref{nobox_ar}
shows that the shape of model $2'$ at $r_{40\%}$ has not significantly change after $60~T_{\rm simu}$. The values of $p$ and $q$ slowly decline
with increasing enclosed mass, but rise again at $r_{80\%}$. At $r_{90\%}$, the axial ratios are $\at~:~\bt~:~\ct=1.00~:~0.96~:~0.66$. The
instability of model $2'$ is driven by non-classified orbits which redistribute to higher energies. The axial ratio of model $3'$ at $r_{\rm 90\%}$ declines to
$1.0~:~0.64~:~0.52$ at $60~T_{\rm simu}$. Compared to the results for model $3$, this indicates that the shape of model $3'$ becomes more prolate. For model $4'$,
the axial ratios as a function of enclosed mass are very similar to that of model $4$. Since the fraction of box orbits in model $4$ is less
than $10\%$, removing them does not substantially affect the evolution of the model. Therefore, we can infer that non-classified orbits play an important role
for the instability in this case.

\subsection{Box orbits and non-classified orbits removed}
\begin{table}
\begin{center}\vskip 0.00cm
\caption{Parameters of the new orbital libraries after removing box and non-classified orbits. The number of total orbits is denoted by $N_{\rm orbit}''$,
$\delta_{\rm smooth}$ is the self-consistency parameter, and $N_{\rm >0,~smooth}''$ is the number of orbits with non-zero weights. The values of the virial
ratio, $-2K/W$, and of the radial stability parameter, $\xi$, are also listed.}
\begin{tabular}{llllc}
\\
\hline
Model  & $2''$ & $3''$& $4''$  \\
\hline
$N_{\rm orbit}''$ &  12000 & 11979 & 11297 \\
$\delta_{\rm smooth}''$ &  $ 0.087$ & $0.087$ & $ 0.108 $\\
$N_{\rm >0,~smooth}''$ &  6077 & 5274 & 820  \\
$-2K/W$ &  1.01 &1.01  &1.01  \\
$\xi$   &  0.98 & 0.83 & 0.98 \\
\hline
\end{tabular}
\label{tab-looponly}
\end{center}
\end{table}

\begin{figure*}{}
\begin{center}
\resizebox{14.5cm}{!}{\includegraphics{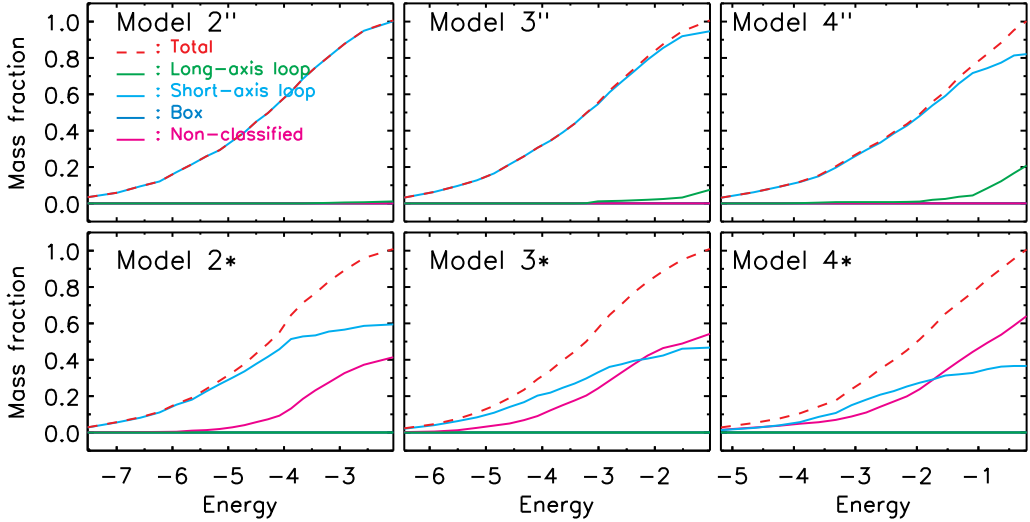}}\vskip 0.5cm
\makeatletter\def\@captype{figure}\makeatother
\caption{The upper panels illustrate the orbital structure of models $2''$--$4''$ computed with Eq. \ref{smoothnnls} after removing box and
non-classified orbits from the orbital library. The lower panels show the orbital structure of models $2*$--$4*$ which have been obtained in
the same way after removing the box and long-axis loop orbits.
}
\label{looponly_of}
\end{center}
\end{figure*}
\begin{figure}{}
\begin{center}
\resizebox{8.5cm}{!}{\includegraphics{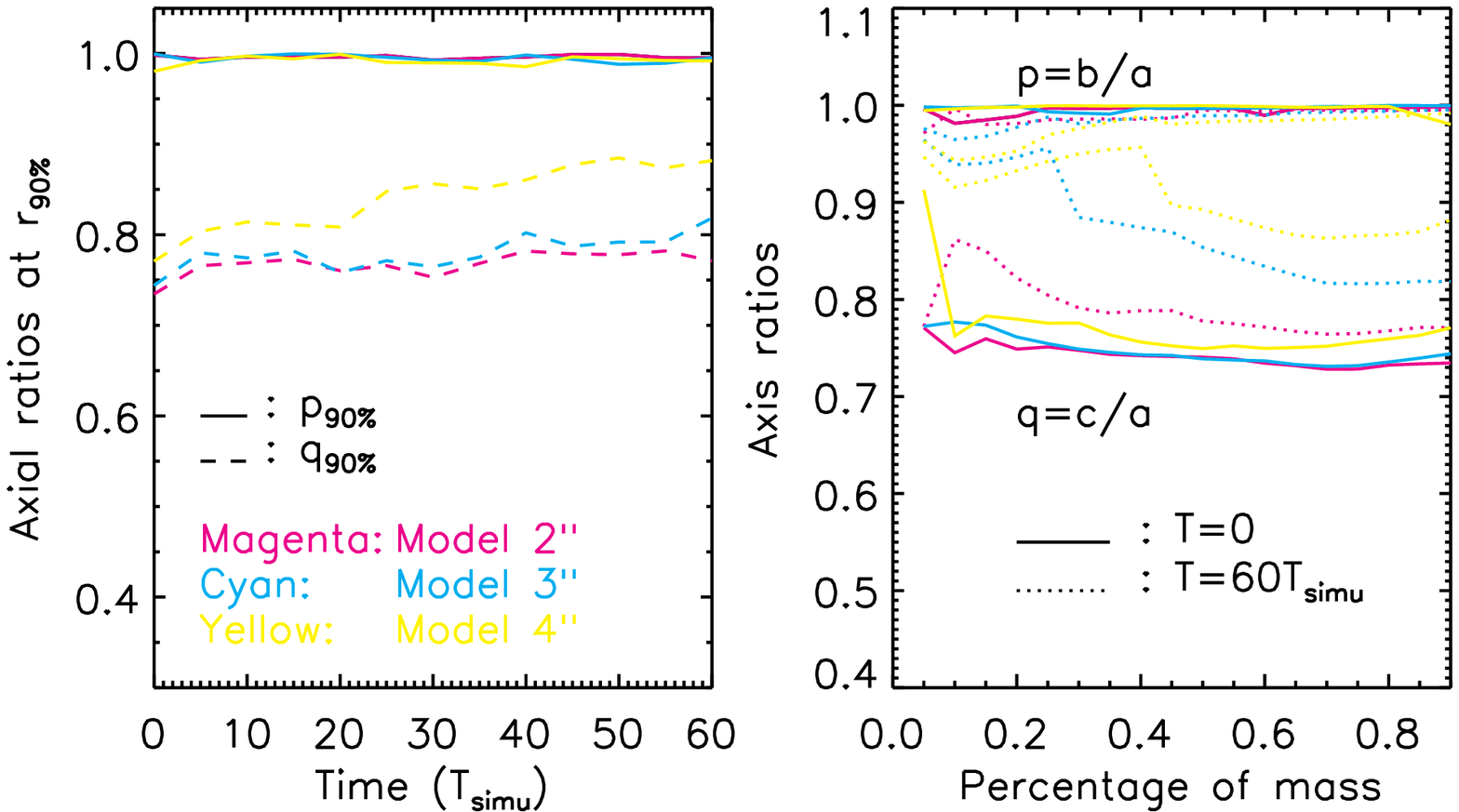}}\vskip 0.5cm
\resizebox{8.5cm}{!}{\includegraphics{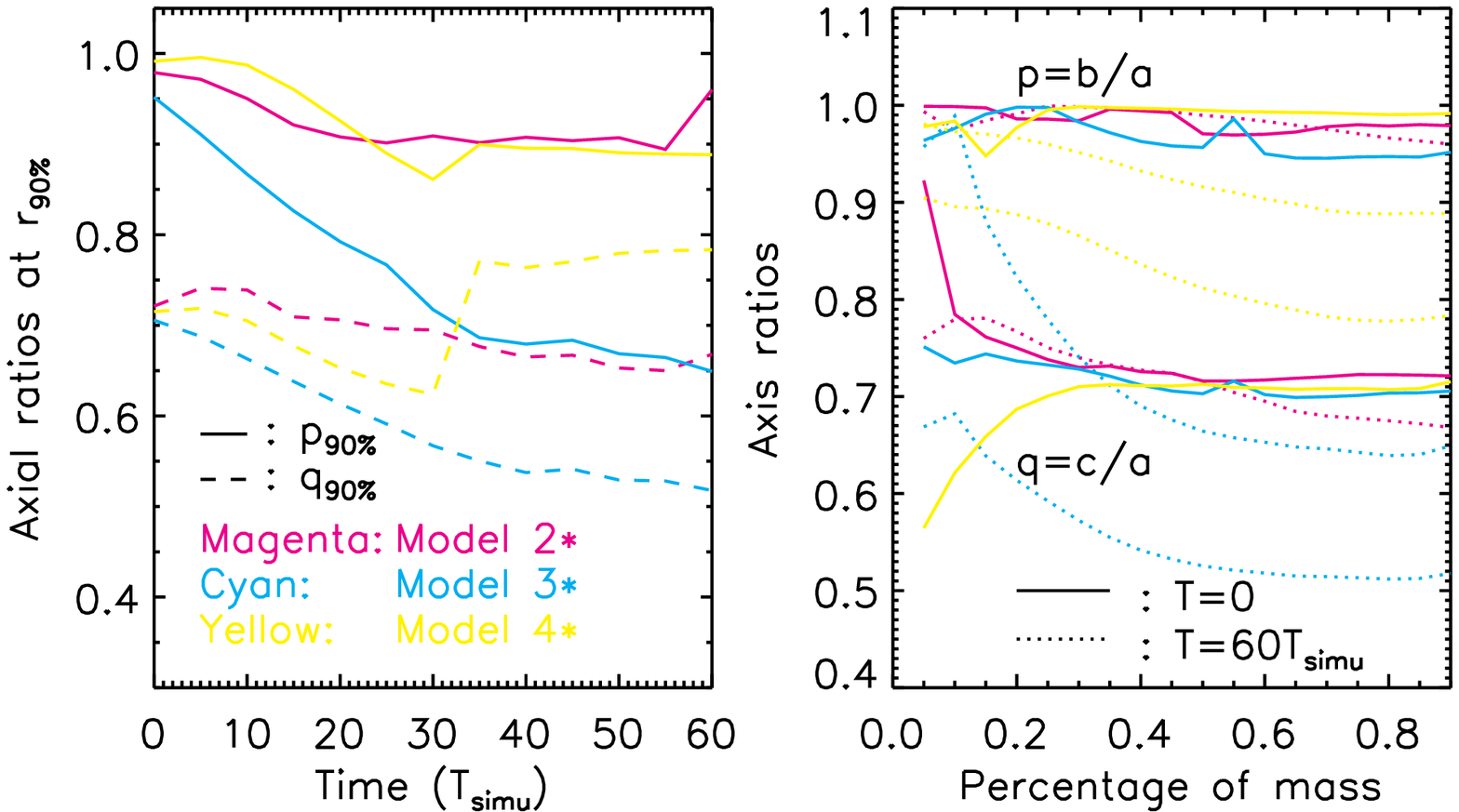}}\vskip 0.5cm
\makeatletter\def\@captype{figure}\makeatother
\caption{The upper panels show the time evolution of axial ratios at $r_{90\%}$ (left) and axial ratios as a function of enclosed masses (right)
for models $2''$ (magenta), $3''$ (cyan),  and $4''$ (yellow). The line types are defined as in Fig. \ref{nobox_ar}. The lower panels illustrate
the results as the upper ones, but not for models $2*$ (magenta), $3*$ (cyan), and $4*$ (yellow).  
}
\label{noboxnox_ar}
\end{center}
\end{figure}

Non-classified orbits are defined as low angular momentum orbits in which the velocity cannot be restricted to a rectangular velocity box (Eq. \ref{box}).
These orbits behave stochastically in grid space. In order to characterise the origin of the instability in the models $2'$--$4'$, we perform a further
test by removing all non-classified orbits from their orbital library. The such obtained models are mostly comprised of short-axis loop orbits and labelled
as models $2'',~3''$ and $4''$. Their orbital structure is illustrated in the upper panels of Fig. \ref{looponly_of}. The models exhibit a small fraction
of long-axis loop orbits and their amount grows with increasing external field strength. It is known that long-axis loop orbits are illegal orbits in both
Newtonian axisymmetric models and isolated axisymmetric MOND models. Their existence might cause additional instability, which we will investigate further
below. The resulting self-consistency parameters $\delta$ are listed in Table \ref{tab-looponly}. In all cases, $\delta\sim 0.1$ which is significantly larger
than for models $2$--$4$ and $2'$--$4'$. This implies that the $N$-body ICs sampled from these Schwarzschild models are not in equilibrium. Therefore, non-classified
orbits are a necessary ingredient for the self-consistency of the models $2$--$4$. The stability of these models is again examined using $N$-body runs as described
in Sec. \ref{sec-nobox}. The estimated axial ratios at $r_{90\%}$ as a function of time and as a function of the enclosed mass at $T=0$ and $T=60~T_{\rm simu}$ are
illustrated in the upper panels of Fig. \ref{noboxnox_ar}. We find that the models are still axisymmetric after $60~T_{\rm simu}$. No triaxial configurations emerge
during the $N$-body experiments, which agrees with the small $\xi$-values estimated for these models. Ignoring the innermost particles within
$r_{20\%}$, the axial ratios of model $2''$ at $60~T_{\rm simu}$ turn out very similar to that of model $1'$, indicating that by removing box and the non-classified
orbits, axisymmetric models embedded in weak external fields evolve almost like isolated ones. Nevertheless, the models become eventually rounder, where the effect
is more prominent for stronger external fields. This might be a result of the model's considerable departure from self-consistency and should be further explored in
future work. Our results suggest that non-classified orbits with low angular momentum likely play an important role for the instability in the original models $3$
and $4$ discussed in Sec. \ref{nbody}. 

\subsection{Box orbits and long-axis loop orbits removed}
\begin{table}
\begin{center}\vskip 0.00cm
\caption{Parameters of the orbital libraries after removing box and long-axis loop orbits. The number of total orbits is denoted by $N_{\rm orbit}*$,
$\delta_{\rm smooth}$ is the self-consistency parameter, and $N_{\rm >0,~smooth}*$ is the number of orbits with non-zero weights. The values of the
virial ratio, $-2K/W$, and of the radial stability parameter, $\xi$, are also listed.}
\begin{tabular}{llllc}
\\
\hline
Model  & $2*$ & $3*$& $4*$  \\
\hline
$N_{\rm orbit}*$ & 19097  & 20381 & 21079 \\
$\delta_{\rm smooth}*$ &  $ 0.053$ & $0.046$ & $ 0.059 $\\
$N_{\rm >0,~smooth}*$ & 1502  & 1570 & 1385  \\
$-2K/W$ &  1.01 &1.01  &1.00  \\
$\xi$   &  1.58 & 1.81 & 1.80 \\
\hline
\end{tabular}
\label{tab-noboxnoxloop}
\end{center}
\end{table}

As mentioned before, external fields break the symmetry of the potential and introduce long-axis loop orbits. However, the density distribution is still
axisymmetric, which conflicts with the presence of long-axis loop orbits. One question that naturally arises is the following: will these long-axis loop
orbits import instability? To address this question, we compare models $2'$--$4'$ to another set of models in which these illegal orbits are removed. The
latter are labelled as models $2*,~3*$ and $4*$, and their orbital structure and evolution is analysed following the same procedure as in Sec. \ref{sec-nobox}.
The orbital structure of the models $2*$--$4*$ is shown in the lower panel of Fig. \ref{looponly_of}. These additional models are comprised of only short-axis
loop and non-classified orbits. As can read from Table \ref{tab-noboxnoxloop}, all models yield self-consistency parameters $\delta \approx 0.05$, which is
close to the corresponding values for models $2'$--$4'$. If long-axis loop orbits introduced another source of instability, the inferred axial ratios of models
$2*$--$4*$ should differ from that of the models $2'$--$4'$ where only box orbits have been removed. The axial ratios of models $2*$--$4*$ are illustrated in
the lower panel of Fig. \ref{noboxnox_ar}. The time evolution of axial ratios at $r_{90\%}$ (lower left panel) is quite similar to that found in models $2'$--$4'$.
Moreover, the axial ratios as a function of enclosed mass at $T=0$ and $T=60~T_{\rm simu}$ (lower right panel) are not significantly altered relative to that
found for the models $2'$--$4'$. Consequently, long-axis loop orbits do not contribute additional instability to the models embedded in external fields. There are, however, small differences in the final configurations between the models $2'$--$4'$ and $2*$--$4*$. For instance, the ratio $p$
for model $2*$ declines mildly (and smoothly) from $r_{50\%}$ to $p\approx 0.96$ at $r_{90\%}$ whereas that of model $2'$ drops more quickly from $r_{40\%}$ and
suddenly increases at $r_{80\%}$. Although the orbital structures of the models $2'$ and $2*$ are nearly identical, there is an $\approx 1\%$ fraction of long-axis loop orbits in model $2'$ and, accordingly, the fraction of non-classified orbits is slightly larger (by about $\approx 0.26\%$). The long-axis loop orbits are mainly distributed at larger radii where about $80\%$ and more of the total mass is enclosed. This is likely responsible for the sudden rise at $r_{80\%}$, i.e., a better agreement with the axial ratios of the original profiles. The slight changes in the orbital structure and non-classified orbit fractions are probably the main reasons for the observed differences between the two models. 
For the models $4*$ and $4'$, the changes in the orbital
structure are more pronounced, with a $10\%$ fraction of long-axis loop orbits in model $4'$ and an increased $\approx ~10\%$ fraction of non-classified orbits in model $4*$. While the shape of model $4'$ at $r_{50\%}$ is more prolate at
$T=60~T_{\rm simu}$, that of model $4*$ turns out more triaxial. The difference between the models $4'$ and $4*$ is due to the different amounts of non-classified orbits. 

To summarise the results of this section, we conclude that both box and non-classified orbits result in instability for isolated models and for models that are
embedded in external fields.

\section{The effects of an external field shock }
The models $1$--$5$ have been constructed in (quasi-) equilibrium from the Schwarzschild models. This is the first attempt to adopt Schwarzschild's technique
for studying models embedded in external fields. It is therefore interesting to compare an equilibrium model embedded in an external field to an isolated model
which suffers from a sudden perturbation due to an external field. The latter model is expected to relax to a new equilibrium state within the applied external
field. Let us define a dynamical time scale for the model,
\beq\label{tdyn}
t_{\rm dyn} \equiv \frac{r_{50\%}}{v_{\rm circ,r50\%}} =r_{50\%}(0.5\times GMa_0)^{-1/4}\approx 3~T_{\rm simu}, \eeq
where $r_{50\%}$ is the half-mass radius. A MOND system revirialising in a new environment has been studied by \citet{Wu_Kroupa2013}, where the system is initially
embedded in a strong external field and the considered in isolation. The system fully revirialises within $5~\tdyn$ defined at the Plummer radius for a Plummer sphere.
Here we are interested in the inverse process. 

\begin{figure}{}
\begin{center}
\resizebox{8.5cm}{!}{\includegraphics[angle=0]{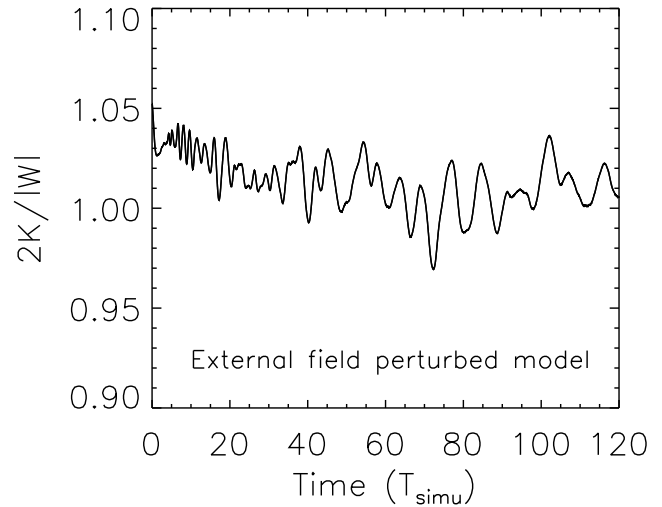}}\vskip 0.5cm
\makeatletter\def\@captype{figure}\makeatother
\caption{The virial ratio of the model $1$ perturbed by an external field of $1~a_0$.}\label{efvir}
\end{center}
\end{figure}
 
We apply an external field with an amplitude of $1~a_0$ to model $1$, and the direction of the external field is assumed to be the same as for model $4$.
The new system is then evolved using the NMODY code. The time evolution of the virial ratio is presented in Fig. \ref{efvir}. The system appears hotter
than the equilibrium state at $T=0$ since the potential becomes shallower in the strong external field, and the virial ratio is about $1.05$ initially.
The system relaxes to a new equilibrium state within about $60 ~T_{\rm simu} \approx 20~ \tdyn$, after which the virial ratio oscillates around unity,
with an amplitude less than $4\%$. 
Although the external field is as strong as $1~a_0$, the most impacted region is at larger radii where $r>8\kpc$
(see the deviation of the circular orbital time between models $1$ and $4$ in Fig. \ref{tcir}). The circular orbital time for model $1$ at $8\kpc$, where there is over $80\%$ of the total mass enclosed, is $\approx 0.3~ Gyr \approx 60~T_{\rm simu} \approx 20~ \tdyn$. The crossing time for the furthest particles in model $1$ is approximately $50~T_{\rm simu}$, which is comparable to the revirialisation time. It explains why the revirialisation time is longer than that observed in \citet{Wu_Kroupa2013}. Further, to be fully relaxed to and further test the stability of the new equilibrium state, the system has been evolved for another $60~T_{\rm simu}$.

\begin{figure}{}
\begin{center}
\resizebox{8.5cm}{!}{\includegraphics{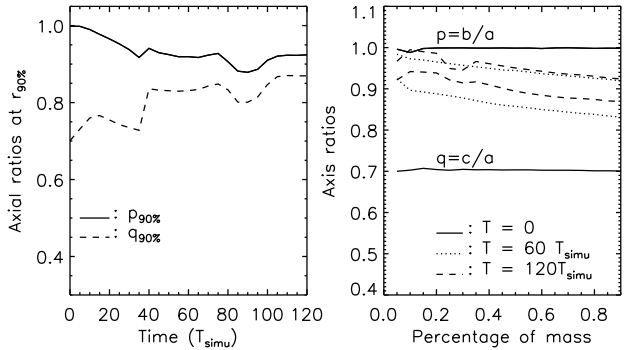}}\vskip 0.5cm
\makeatletter\def\@captype{figure}\makeatother
\caption{\textbf{Left panel}: Time evolution of the principle axes at $r_{90\%}$ for model $1$ perturbed by a strong external field.
The solid, dotted and dashed lines denote the different components $\at$, $\bt$, and $\ct$, respectively. \textbf{Right panel}: Axial
ratios at three evolutionary times as a function of enclosed mass. The upper set of curves refers to $p=\bt /\at$ and the lower set to
$q=\ct /\at$. The solid, dotted and dashed curves represent the initial ($T=0$), intermediate ($T=60~T_{\rm simu}$), and final ($T=120~T_{\rm simu}$)
axial ratios.}
\label{efar}
\end{center}
\end{figure}

The axial ratios for particles within $r_{90\%}$ for the shocked model are shown in the left panel of Fig. \ref{efar}. The values of $\pr$ and $\qr$
evolve within the first $40~T_{\rm simu}$, where $\pr$ decreases from $1.00$ to $0.94$, $\qr$ increases from $0.70$ to $0.84$, and the model becomes
prolate, with axial ratios $1.00~:~0.94~:~0.84$. The global shape then remains stable for the next $40~T_{\rm simu}$. At $T=80~T_{\rm simu}$, there
is an inflection for both $\pr$ and $\qr$, and the model becomes even more prolate, with an axial ratio as low as $1.00~:~0.88~:~0.80$ at $T\approx 90 T_{\rm simu}$.
The ratios return to $1.00~:~0.92~:~0.87$ at $T\approx 105 T_{\rm simu}$, and the system evolves into a triaxial configuration. The behaviour within
$80$--$105~T_{\rm simu}$ implies an instability in the model's evolution.

The right panel of Fig. \ref{efar} shows the model's axial ratios at $T=0$ (solid curves), after revirialisation at $T=60~T_{\rm simu}$ (dotted curves),
and at $T=120~T_{\rm simu}$ (dashed curves). The oblate axisymmetric model turns prolate into a prolate one after revirialisation in the external field
at all Lagrangian radii. After reaching a new equilibrium state, the model further exhibits instability and becomes triaxial after another $60~T_{\rm simu}$
with axial ratios $1.00~:~0.92~:~0.87$ at $r_{90\%}$. The resulting density configuration is less prolate than in model $4$, suggesting that such shocked
external field models are rounder than their counterparts which are initially constructed in (quasi-) equilibrium.

\begin{figure}{}
\begin{center}
\resizebox{8.5cm}{!}{\includegraphics{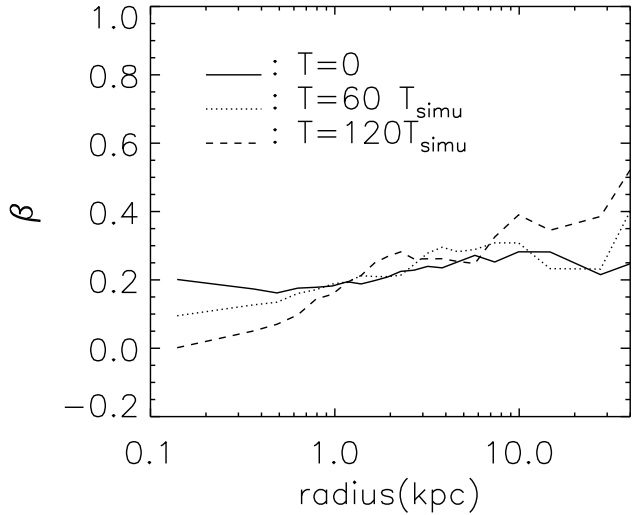}}\vskip 0.5cm
\makeatletter\def\@captype{figure}\makeatother
\caption{Anisotropy parameter $\beta(r)$ for the external field shocked model. The solid, dotted and dashed curves represent the initial ($T=0$),
intermediate ($T=60~T_{\rm simu}$) and final ($T=120~T_{\rm simu}$) states.}
\label{efbeta}
\end{center}
\end{figure}

As the external field leads to a shallower potential, the model particles are less bound than in the unperturbed case. Hence, the particles can move along
more radial orbits, which is especially true for particles in the outer region. We thus expect an increase of $\beta(r)$ in the model's outer region after
revirialisation and further evolution. This is confirmed in Fig. \ref{efbeta}. The anisotropy parameter $\beta(r)$ grows up to $0.4$ at $T=60~T_{\rm simu}$
and to $0.5$ at the end of the simulation. Within $1~\kpc$, $\beta(r)$ decreases and eventually reaches values close to zero. The final $\beta(r)$-profile
within $1~\kpc$ is very similar to that of model $1$ because the external field does not significantly affect the orbits of the innermost particles.

For an adiabatic scenario in which the external field is gradually increased, we do not expect any prominent differences from the shocked model. Considering
the opposite setup, i.e. a system embedded in a strong external field that is suddenly isolated, \citet{Wu_Kroupa2013} have not found any distinct features
between the behaviour in the shocked and adiabatic situations, although the evolution of $\beta(r)$ is milder in the adiabatic case. Investigating this issue
in more detail, however, is beyond the scope of this paper.

\section{Lopsidedness}
\begin{figure}{}
\begin{center}
\resizebox{8.5cm}{!}{\includegraphics{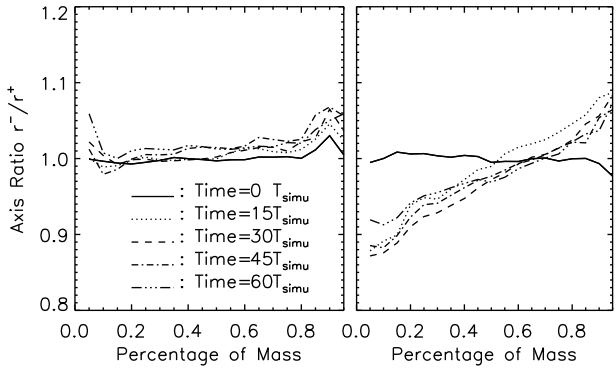}}\vskip 0.5cm
\makeatletter\def\@captype{figure}\makeatother
\caption{The evolution of lopsided shapes in terms of $r^-:r^+$, assuming model $4$ \textbf{(left panel)} and model $5$ \textbf{(right panel)}.
Different line types refer to different simulation snapshots.
}\label{lopsidedness2}
\end{center}
\end{figure}

In Fig.~\ref{isopot}, we have illustrated the lopsided shape of the potential for an axisymmetric input density in an external field. The difference between the potential
and the density distribution will lead to an additional relaxation of the model, and the resulting lopsided morphologies, which are further characterised by an offset between
the density peak and the centre of mass, have previously been studied in \citet{Wu_etal2010}. Nevertheless, it is unknown whether the found lopsided shape is a stable feature.
By defining the axis ratio
\beq
{r^{-}}:{r^{+}} = \sqrt{\sum_i m_ix_{i,-}^2/\sum_i m_i}:\sqrt{\sum_j m_jx_{j,+}^2/\sum_j m_j},
\eeq
where $\pm$ denotes the upper/lower half space with respect to the $x$-axis, we can study the the shape of mass distribution at different times during the evolution. In both
panels of Fig.~\ref{lopsidedness2}, we find that the axis ratios of the ICs are symmetric within $r\sim r_{90\%}$. After the evolution, the models become lopsided at large
radii. The lopsided feature becomes stable within $30~T_{\rm simu}$ for both models $4$ and $5$. The resulting values of $r^-:r^+$ for model $4$ and model $5$ are around
$1.05$ and $1.07$, respectively. Both models $4$ and $5$ correspond to very bright ellipticals, and such lopsided features are expected to be more prominent for low-luminosity
ellipticals within similar external fields.

This suggests that one cannot find perfectly symmetric galaxies in the centres of clusters within the framework of MOND. External fields will distort the
original symmetry when the gravitational field strength starts to become comparable to that of the internal one, which is especially true for the outer
parts of a galaxy.

\begin{figure*}{}
\begin{center}
\resizebox{14cm}{!}{\includegraphics{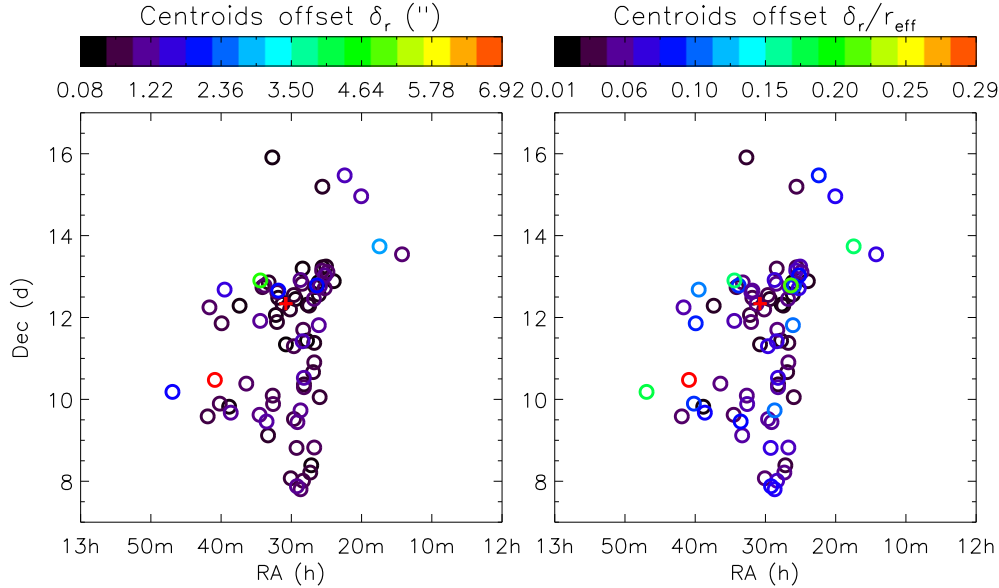}} 
\caption[Centroids offset of dwarf ellipticals in the Virgo cluster]{The offsets of nuclei in dwarf elliptical galaxies inside the Virgo cluster: The coloured empty circles
are the 78 dwarf ellipticals observed on the sky \citep{Binggleli_etal2000}, and the red cross denotes the position of Virgo's centre. The colours on the \textbf{left panel}
illustrate the centroids' offsets $\delta_r$ in units of arcsec ($1^{\prime\prime} \sim 100~\pc$), those on the \textbf{right panel} relate to the relative offsets
$\delta_{r}/r_{\rm eff}$, where $r_{\rm eff}$ is the effective radius of a dwarf galaxy.}
\label{virgo}
\end{center}
\end{figure*}

The observations of 78 ``nucleated'' dwarf elliptical galaxies in Virgo \citep{Binggleli_etal2000} show that $20\%$ of the sample are drastically lopsided.
The associated typical centroid offset of these galaxies is about $100~\pc$ (assuming that Virgo is at a distance of $20~\mpc$). The other dwarfs in the
sample appear also lopsided, but less dramatic (see Fig.~\ref{virgo}). In the context of MOND, the magnitude of the offset associated with the lopsided shape
should correlate with the strength of the external field. For galaxies far from the cluster centre, the gravitational environment becomes weaker and one
expects smaller offsets. Figure~\ref{virgo} shows the absolute offsets of nuclei (\textbf{left panel}, in units of arc-second) and the relative offsets
${\delta_{r}/r_{\rm eff}}$ (\textbf{right panel}). Most of these dwarf galaxies exhibit offsets that are larger than about 5\% of their effective radii $r_{\rm eff}$.

To assess whether observations of this kind could be used to constrain the MOND paradigm, we need to investigate how a three-dimensional correlation
between offsets and distances from the cluster centre would appear when looking at the corresponding projected quantities. For this purpose, we have
considered a simple numerical experiment. Assuming a linear relation between $\delta_{r}$ and the external field such that $\delta_{r}(a_{0}) = 200~\pc$
\citep{Wu_etal2010}, we created several random realizations of $\sim 100$ galaxies within a Virgo-like potential, adopting the NFW profile of \citet{McLaughlin1999}.
Ignoring any further errors or physical effects, we then simply compare estimates of the linear correlation coefficient $Q$ between offsets and radii before
and after projecting along the line of sight. As a result, we find that the correlation is significantly reduced, with $-Q$ dropping from values of $0.9$--$0.95$
down to $0.4$--$0.45$. Given additional uncertainties such as the presence of tidal fields, selection biases, and other degenerate effects \citep{Binggleli_etal2000},
we conclude that the current data are not enough to falsify or support MOND. Future high-resolution observations for nearby clusters, however, might improve this situation.

\section{Conclusions and discussions}
\label{concl}
Following Schwarzschild's approach, we constructed (quasi-) equilibrium models for galaxies with a central cusp embedded into uniform external fields within
the framework of MOND. For these models, we performed instability tests and kinematic analyses by means of $N$-body simulations which operate on a spherical
grid \citep{NMODY}.

When applying Schwarzschild's method to galaxies in external fields, the internal potential is distorted and the models are not exactly self-consistent with
respect to the original analytic density profile. This leads to non-equilibrium initial conditions which relax to a dynamic equilibrium within a few simulation
times (see Fig. \ref{virial}). Since the overall residual is only at the level of a few percent, the deviation from the equilibrium state is rather minor. For
comparison, the isolated models discussed in \citet{stability} are found in a perfect equilibrium.

Interpreting the results of our $N$-body simulations, we conclude that galaxy models within external fields appear unstable over at most $\sim 30$ dynamical
times. The shapes of these systems evolve due to instability. The density profiles along the major axes of the strong external field models clearly change after
$30~T_{\rm simu}$, while those of other models remain close to the initial Hernquist profiles. In contrast to the stable isolated triaxial systems considered in
\citet{stability}, the isolated axisymmetric model is also unstable due to illegal box and non-classified orbits with low angular momentum.
It evolves towards a triaxial model with axial ratio close to $1.00~:~0.90~:~0.71$ within $r_{90\%}$. If box orbits are removed from the orbital library at the
cost of self-consistency, the orbital selection procedure suppresses non-classified orbits and the isolated axisymmetric model becomes stable.

For equilibrium systems embedded in weak and intermediate external fields, the final evolved shapes are similar to that of the isolated axisymmetric model. The
presence of a strong external field yields a more prolate shape for axisymmetric models based on the same initial density profile. Similarly, the final shape of
the triaxial model we considered within a strong external field turns out slightly more prolate. For these models, both box and non-classified orbits contribute
to the instability. Long-axis loop orbits, which appear in the asymmetric potential due to the external field, however, do not. Further evidence for the instability of all models (including the isolated case) is provided by the temporal evolution of $\sigma_r(r)$ and $\beta(r)$, especially in their inner regions.

We have also studied the case of an isolated axisymmetric model which is perturbed by a strong external field. Shocked by the external field, this model revirialised
to a new equilibrium state after $60~T_{\rm simu} \approx 20~\tdyn$, and then evolved through instability during the following $45~T_{\rm simu}$. The final shape of
the model is also prolate, but rounder than that of the corresponding self-consistent model in the same external field. The shocked case further shows an increase of
radial anisotropy in the outer region.

While there is observational evidence for lopsided dwarf galaxies in Virgo, it is inconclusive whether this could be clearly linked to external field effects in MOND
for future datasets. Since the MOND lopsidedness appears on the system's outskirts, accounting just for a small fraction of the total mass, the effect is expected to
be rather small. 

Here we used the simple $\mu$-function defined by Eq.~\ref{simplemu} for all considered models. Compared to the standard form used in \cite{Milgrom1983c},
it leads to a more gradual transition from the deep MOND regime to Newtonian gravity. If the standard $\mu$-function was adopted, the models would be subject to weaker
MOND effects in the intermediate field regions where gravity is comparable to $a_0$ (i.e., at the radius $\approx~8~\kpc$ for models $1-4$ and $\approx~11~\kpc$ for model
$5$; see Fig. \ref{tcir}). The use of different interpolating function in the Schwarzschild modelling will result in a change of the orbital structure. Since a MOND
spherical system with radial anisotropy is more stable than a pure Newtonian model with exactly the same density distribution \citep{Nipoti_etal2011}, one may expect
more severe instability in MOND models based on the standard $\mu$-function if the fraction of box and non-classified orbits with low angular momentum is comparable
to what we have found in our analysis.

It is interesting to consider the consequences of an evolving external field associated with the formation galaxy clusters or galaxy groups. As galaxy clusters grow in
density, the external field amplitude grows as well, which suggests an increase of radially anisotropic triaxial cluster galaxies as a function of decreasing redshift.
Such a trend might be testable through detailed statistics of substructure shapes from lensing data of galaxy clusters. Considering the local Universe, one surprising
prediction of MOND is that the Andromeda galaxy should have had a fly-by within 30 kpc of the Milky Way at about 7 Gyrs ago (cf. \citealt{Zhao_etal2013,Banik_Zhao2016,
Banik_Zhao2017}). In this scenario, the two galaxies were drawn together by the greatly enhanced mutual attraction in MOND, and flung out efficiently with very little
dynamical friction as shown in preliminary $N$-body and hydro simulations (Banik, Renaud, private communications, 2017). From the view of the external field problem,
one would expect that at the time of fly-by, both galaxies suffered an external field shock with a peak value of $g_{\rm ext}\approx (200\kms)^2/30\kpc\approx 0.3a_0$.
This external field might have had a stronger effect on the smaller system, i.e. our Galaxy, and could have triggered the formation of the triaxial bar in the centre.

It is also worth cautioning that the external field effect is theory-dependent. For instance, the superfluid theory \citep{Khoury2015,Khoury2016,Berezhiani_Khoury2015,
Berezhiani_Khoury2016} is one of the recent attempts to conciliate MOND with the cold dark matter phenomenology in galaxy clusters. In such theories, the MOND effect
is achieved only on scales of 50 kpc inside a small superfluid core of the cluster, and our predictions would {\it not} apply to galaxies located in the bulk or the
outer parts of clusters.

\section{Acknowledgments}
We thank the anonymous referee for careful and helpful suggestions and comments to the earlier version of the manuscript. The authors thank Luca Ciotti, Pasquale Londrillo, Carlo Nipoti for generously sharing their code. The main body of the work has been accomplished at the University of Science and Technology of China and is supported by the NSFC grants 11503025, 11421303, and by ``the Fundamental Research Funds for the Central Universities''. An early stage of this work has been performed under the Project HPC-EUROPA (211437), with the support of the European Community - Research Infrastructure Action under the FP8 ``Structuring the European Research Area'' Programme. XW thanks Simon Portegies Zwart for his warm hospitality at Leiden University. XW is grateful to James Binney and Keith Horne for their valuable comments and suggestions at an early stage of the project. XW thanks the support from ``Hundred Talents Project of Anhui Province''.
Part of this project was carried out during the PhD programme of XW. XW, MF, and HZ acknowledge partial support from the Scottish Universities Physics Alliance (SUPA). YGW acknowledges support of the 973 Program (No. 2014CB845703), and the NSFC grants 11390372, 1163004. The research of MF was supported by the I-CORE Program of the Planning and Budgeting Committee, THE ISRAEL SCIENCE FOUNDATION (grants No. 1829/12 and No. 203/09), and the Asher Space Research Institute. MF acknowledges support through a fellowship from the Minerva Foundation.

\bibliographystyle{aasjournal}
\bibliography{lopside}

\begin{appendix} 
\renewcommand{\appendixname}{Appendix~\Alph{section}}
\section{A. Integration of orbits}\label{OrbIntegration}
\subsection{A.1. Symmetry and grid segmentation of the models}
We use an orbital integration code which adopts a $7/8$ order Runge-Kutta method \citep{Fehlberg1968} to ensure sufficient accuracy of the orbits
\citep{triaxial}. The way we divide the spatial grid is exactly the same as in \cite{triaxial} (also see \citealt{stability} for numerical details).
Here we consider a total of $1008$ equal mass cells. The outermost $48$ cells are not taken into account because their boundaries extend to infinity.
Since the density evolves as $\rho \sim r^{-4}$ at large radii, the orbits in this sector should contribute much less than all other orbits, and thus
they are negligible.\footnote{The work of \cite{triaxial} ignored the two outermost sectors, thus there were only $912$ cells in the models. Here we
discard only the last sector because the galactic outskirts might extend to larger radii in an external field. In addition, our available hardware has
significantly improved, allowing us to consider these very diffuse outskirts.}

In a previous study, \citet{triaxial} and \citet{stability} considered isolated triaxial models. Due to the three-folded symmetry, they took only the
first octant into account, and then symmetrised the $O_{ij}$ by reflecting all orbits at the octant's boundaries, i.e. the $x$-$y$, $x$-$z$, and $y$-$z$
planes. We use the same approach for the isolated model $1$.

As the presence of an external field breaks the symmetry of the resulting potentials, we cannot simply reflect the orbits at these planes anymore. For
model $5$, the external field points into negative $x$-direction, and thus orbits starting from the positive $x$-semi-space are not the mirror orbits of
those starting from the negative $x$-semi-space, i.e. orbits on different sides of the $y$-$z$ plane behave differently. However, the orbits will still
have a two-folded symmetry with respect to the $x$-$z$ and $x$-$y$ planes. Hence we used the first ($|x|$,$|y|$,$|z|$) and second ($-|x|$,$|y|$,$|z|$)
octants to calculate orbits for model $5$. The total number of cells in this case is set to $960\times 2 = 1920$, where the outermost $96$ cells are
again excluded.

When the external field is pointing into an arbitrary direction, the symmetry of the model is further reduced. Switching to an axisymmetric model and
forcing the external field to point into a direction perpendicular to the symmetry axis (see any of the models $2$--$4$ in Table~\ref{mass}), however,
we can always find a coordinate frame such that the external field is parallel to the $x$-axis. As the potential will be distorted along the external
field direction, we lose the symmetry along the $x$-axis in addition to the one along the $z$-axis. However, the potential will still be symmetric with
respect to the $x$-$z$ plane, which can be exploited to simplify the numerical computation. We fold the system at the $x$-$z$ plane and consider four
octants, the first ($|x|$,$|y|$,$|z|$), second ($-|x|$,$|y|$,$|z|$), fifth ($|x|$,$|y|$,$-|z|$) and sixth ($-|x|$,$|y|$,$-|z|$). We calculate orbits by
reflecting them at the $x$-$z$ plane. In this case, we have $960\times 4=3840$ mass cells (the outermost 192 cells are excluded) dividing the semi-space.
Note that if the system is triaxial and the external field direction does not coincide with one of the axes, we need to consider all eight octants, using
$960\times8=7680$ cells. Since this is computationally expensive, we limit ourselves to axisymmetric systems (models $2$--$4$) and a single triaxial
system where the external field is anti-parallel to the $x$-direction.

\subsubsection{A.2. Initial conditions for orbits and orbital classification}
Following \cite{Schwarzschild1993} and \cite{Merritt_Fridman1996}, we set up the ICs for the orbit library. The numerical details of this procedure can
be found in \citet{triaxial,stability}. Here we only give a brief description. There are two sets of ICs for the starting points:
\begin{enumerate}
\item Stationary starting orbits from $20$ equal potential surfaces, corresponding to the $20$ nodes of equal mass sectors on the $x$-axis. These are
freely falling into the system's internal potential with their initial velocities set to zero. Since these orbits do not carry any angular momentum
initially, they may cross the centre of the system and change the sign of their angular momenta. It is known that stationary initial conditions can
produce box orbits. Defining a velocity box through
\beq
\frac{\max\left (3v_{x}^{2}\right )\cdot\max\left (3v_{y}^{2}\right )\cdot\max\left (3v_{z}^{2}\right )}
{\max\left (v_{x}^{2} + v_{y}^{2} + v_{z}^{2}\right )^{3}} < 1,
\label{box}
\eeq 
we classify orbits as box orbits if they satisfy the above inequality which describes an almost rectangular box inside a maximum-energy sphere in velocity
space. Note that the time-averaged angular momentum is zero.
\item Ejecting orbits starting from the $x$-$z$ plane with initial velocities $(v_{\rm x},v_{\rm y},v_{\rm z})=(0,\sqrt{2(E-\Phi_{\rm int})},0)$. Here $E$
corresponds to the total potential energy of the 20 equal mass sectors along the $x$-axis, and $\Phi_{\rm int}$ is the (smooth) internal potential. Since
most of the ejecting orbits are loop orbits, they cannot enter the system's central part. There are two families of loop orbits since the initial components
$L_{x}$ and $L_{z}$ are not zero. We classify orbits which conserve the sign of their angular momentum around the long axis during the whole simulation
as the long-axis loop orbits. Alternatively, this is expressed in terms of
\beq
\max\left (L_{x}\right )\cdot\min\left (L_{x}\right ) > 0.
\eeq 
Similarly, we define short-axis loop orbits as 
\beq
\max\left (L_{z}\right )\cdot\min\left (L_{z}\right ) > 0.
\eeq 
These orbits cannot move across the centre of the system because their angular momentum at this point is zero. Other types of orbits, which are beyond the
definition of boxes and loops, are referred to as non-classified orbits.
\end{enumerate}

With the above ICs, one can generate most of the orbits in full phase space for a given potential \citep{Schwarzschild1993}. The starting points of the
orbits for model $1$ are the same as in \cite{stability}. For model $5$, we add mirror positions in the second octant, and for the models $2$--$4$, we use
four mirror octants as mentioned above. In Table~\ref{mass}, we summarise the ICs for the five models, where $N_{\rm stationary}$ and $N_{\rm ejecting}$ are
the numbers of the starting points. Note that with a higher number of cells and orbits, both the size of the array $O_{ij}$ and the linear system (Eq.
\ref{linear}) increase, and building the orbit library becomes computationally more expensive as has been specified in Sec \ref{sec-tcir}.

\section{B. Further dynamical studies on the systems' stability}\label{furtherdyn}
\subsection{B.1. Kinetic energy}
\label{section44}
Since the inner and outermost parts of non-isolated models exhibit some radial instability, probably causing a few particles to leave the system during this
phase, we only consider the remaining (more stable) fraction for further study. Since the left panels of Fig. \ref{enclosed} show that the $90\%$ Lagrangian radii for all the models stay constant within $60~T_{\rm simu}$, we shall use the particles within $r_{\rm 90\%}$.

\begin{figure}{}
\begin{center}
\resizebox{8.5cm}{!}{\includegraphics{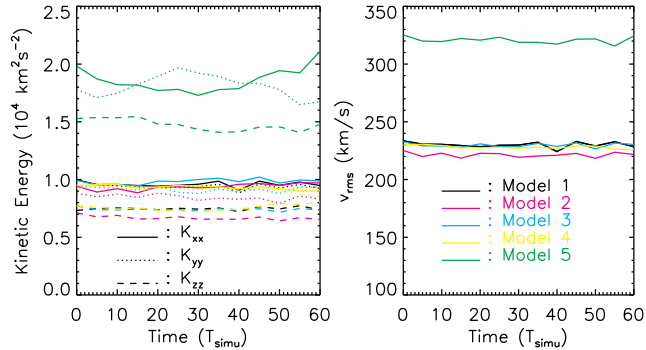}}\vskip 0.5cm
\makeatletter\def\@captype{figure}\makeatother
\caption{\textbf{Left panel}: Time evolution of the kinetic energy tensor components $K_{\rm xx}$, $K_{\rm yy}$, and $K_{\rm zz}$. The colours black, magenta,
cyan, yellow and green represent models $1$--$5$, respectively. The solid, dotted and dashed lines denote the different components. \textbf{Right panel}: Time
evolution of the root-mean-square velocities. The colours are defined as in the \textbf{left panel}. }\label{kinetic}
\end{center}
\end{figure}

As the velocities redistribute substantially in the model $2$--$5$, the kinetic energy of the systems could also change by a significant amount. The \textbf{left panel}
of Fig. \ref{kinetic} illustrates the kinetic energy tensor components $K_{\rm xx}$, $K_{\rm yy}$ and $K_{\rm zz}$ which are given by
\beq\label{kxxef}
K_{\rm xx} = {1\over 2}{ \sum_i m_{i}v_{x}^{2}\over \sum_i m_{i}}
\eeq
for $r<r_{90\%}$ and analogously for the other directions. We highlight that $K_{\rm xx}$, $K_{\rm yy}$ and $K_{\rm zz}$ are computed with respect to the original
$(x,~y,~z)$ coordinate frame instead of the eigenframe considered in Sec. \ref{sec-shape}. Since all particles have equal mass $m_{i} = M/N$, Eq. ~\ref{kxxef}
simplifies to $K_{xx} = 0.5 \langle v_{x}^{2}\rangle$. We find that model $1$ (black lines) does not exhibit any prominent alterations during $60~T_{\rm simu}$.
The components $K_{\rm xx}$, $K_{\rm yy}$, and $K_{\rm zz}$ of model $1$ evolve almost identically. Compared to the initial values, oscillations
in the individual components are smaller than $10\%$ within $60~T_{\rm simu}$.
At the beginning, the kinetic energy components $K_{\rm xx}$ (solid lines) and $K_{\rm yy}$ (dotted line) of model $2$ (magenta), model $3$ (cyan) and model $4$
(yellow) coincide since the ICs are axisymmetric. The component $K_{\rm xx}$ does not particularly change with time for these models. The amplitudes
of oscillations around the original values are less than $6\%$. Although initially having the same density profile, the components $K_{\rm yy}$ of the models $2$
and $3$ evolve quite different from that of model $1$. The component $K_{\rm yy}$ of the models $2$ and $3$ slightly drops from about $20~T_{\rm simu}$. Compared to
their initial values, $K_{\rm yy}$ decreases by about $5\%$ and $13\%$ for model $2$ and $3$, respectively, after $60~T_{\rm simu}$. For model $4$, where the
external field is the strongest, $K_{\rm yy}$ falls from about $50~T_{\rm simu}$ and becomes around $8\%$ smaller than at $T=0$. The component $K_{\rm zz}$ of model
$2$ starts to decrease from $10_{\rm simu}$ and is reduced by about $8\%$ at the end of the simulation; for the models $3$ and $4$, it oscillates around the initial
values with amplitudes smaller than $10\%$. Variations of the kinetic energy components signal model instability. In Sec. \ref{OrbIntegration}, we have seen that the fraction of box orbits in the models $2$--$4$
decreases while the fraction of non-classified orbits increases. The redistribution of kinetic energy for these axisymmetric models might mainly be influenced by instability
due to box orbits.

The situation for model $5$ is more complex. At the beginning of the simulation, we have $K_{\rm xx}>K_{\rm yy}>K_{\rm zz}$ since the model is triaxial. The component
$K_{\rm xx}$ starts to drop whereas $K_{\rm yy}$ rises, and they intersect at $\approx 14~T_{\rm simu}$. The component $K_{\rm yy}$ reaches its maximum value, (up to $\approx 10\%$ relative to its original value) at $\approx 25~T_{\rm simu}$, and $K_{\rm xx}$ becomes minimal, (reduced
by $\approx 13\%$ relative to the original value, at $\approx 30~T_{\rm simu}$. The two components meet again at $\approx 42~T_{\rm simu}$. The component $K_{\rm zz}$
does not considerably evolve, and oscillates around the initial value with an amplitude less than $8\%$.

The \textbf{right panel} of Fig.~\ref{kinetic} shows the time evolution of the root-mean-square velocities $v_{\rm rms}$. For all models, the $v_{\rm rms}$-profiles remain
stable with tiny ($\pm 4\%$) and smooth oscillations. Therefore, the overall kinetic energy is conserved during the simulations. 

\subsection{B.2. Angular momentum}\label{angmom}
\begin{figure}{}
\begin{center}
\resizebox{8.5cm}{!}{\includegraphics{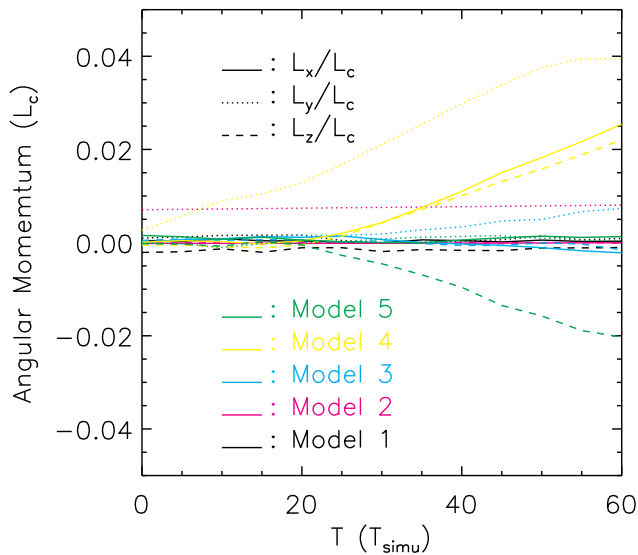}}\vskip 0.0cm
\makeatletter\def\@captype{figure}\makeatother
\caption{Evolution of angular momenta $L_{\rm x}$, $L_{\rm y}$ and $L_{\rm z}$. of models with external fields (model $1$ and $3$). The colours black, magenta,
cyan, yellow and green represent models $1$--$5$, respectively. The solid, dotted and dashed lines denote the different components.}
\label{inertia}
\end{center}
\end{figure}

The external field generally reduces the symmetry of a given model, which could give rise to a self-rotation. Another consequence of the external field is that
--- unlike the case of an isolated system --- the system's (local) angular momentum is typically no longer conserved. The external field introduces additional
torque into the system, and hence we already expect to encounter a variation of this quantity in our simulations. To further investigate these effects, we have
calculated the angular momenta for all models. The results are shown in Fig.~\ref{inertia} which illustrates the unit-mass angular momentum components
$L_{\rm x}$, $L_{\rm y}$ and $L_{\rm z}$ in units of $L_{\rm c}$,
\beq
L_{\rm x} = \frac{1}{M}\sum_{i=1}^{N_p}m_{i}L_{x}^{(i)}
\eeq
and analogously for the other directions, where $N_p$ is the number of particles inside $r_{90\%}$ and $L_{\rm c}$ is defined as the unit-mass angular momentum
with circular velocity $v_{\rm c}$ at a radius $r_{\rm c} = 1$ kpc, $L_{\rm c} = r_{\rm c} v_{\rm c}$. The values of $L_{\rm c}$ are listed in Table~\ref{delta}.
We emphasise that the angular momentum components are computed with respect to the original $(x,~y,~z)$ frame.

A non-zero angular momentum is not expected for model $1$ because it is an axisymmetric model and the system is initially not globally rotating. The black curves
in Fig. \ref{inertia} confirm this. The angular momentum of model $2$ is also conserved during the simulation. Although there is a tiny non-zero value of $L_{\rm y}$,
the values do not change noticeably within $60~T_{\rm simu}$. For the models $3$--$5$, there are interesting evolutions of angular momentum. The component $L_{\rm y}$
of model $3$ starts increasing from $30~T_{\rm simu}$ and has a small value of $0.007~L_{\rm c}$, which is comparable to the constant $L_{\rm y}$ of model $2$. The
increase of angular momentum is caused by the external field. For model $4$, the evolution of angular momentum is more significant. In this case, $L_{\rm y}$ changes
from the beginning of the simulation and grows up to $0.04~L_{\rm c}$ at $60~T_{\rm simu}$. The components $L_{\rm x}$ and $L_{\rm z}$ begin to increase from
$\approx 20~T_{\rm simu}$ and grow in nearly the same fashion, again since the direction of the strong external field is along the diagonal negative $x$-$z$ direction.
We find $L_{\rm x}\approx 0.025~L_{\rm c}$ and $L_{\rm z}\approx 0.022~L_{\rm c}$ at the end of the simulation. The components $L_{\rm x}$ and $L_{\rm y}$ of model
$5$ do not evolve since the external field lies along the $x$-axis and there is no additional torque introduced by the external field. Nevertheless, $L_{\rm z}$ of
model $5$ grows in the opposite direction from $\approx 20~T_{\rm simu}$ and $L_{\rm z} \approx -0.02$ at $60~T_{\rm simu}$. The growth of angular momentum at around
$20-30~T_{\rm simu}$ for models $3$--$5$ might be a result of model instability because of the time coincidence. 
However, the evolution of angular momentum is quite sophisticated and less intuitive as it depends on the assumed direction of the external field, the actual orbit
composition, and the level of numerical noise which increases with simulation time.

Finally, we point out that the observed changes of angular momentum components are relatively small, corresponding to a few $\kpc~\kms$. Consequently, any self-rotation
in the models caused by the external fields must be tiny.

\end{appendix}

\end{document}